 \documentclass[final,5p,times,twocolumn,authoryear]{elsarticle}

\usepackage{amssymb}
\usepackage{amsmath}
\usepackage{bm}
\usepackage{url}
\usepackage{xcolor}
\usepackage{ulem}

\journal{New Astronomy}

\begin{document}
\begin{frontmatter}



\title{Collision frequency between dark matter subhaloes within Milky Way-like galaxies}


\author[a,b,c]{Koki Otaki}
\ead{koki.otaki@uniroma1.it}
\affiliation[a]{organization={Dipartimento di Fisica, ``Sapienza'' Università di Roma},
            addressline={Piazzale Aldo Moro 5},
            city={Rome},
            postcode={I-00185},
            state={},
            country={Italy}}
\affiliation[b]{organization={INAF/Osservatorio Astronomico di Roma}, addressline={Via Frascati 33}, postcode={00040}, city={Monte Porzio Catone},state={}, country={Italy}}
\affiliation[c]{organization={INFN, Sezione Roma1, Dipartimento di Fisica, ``Sapienza'' Università di Roma}, addressline={Piazzale Aldo Moro 2}, postcode={I-00185}, city={Roma},state={}, country={Italy}}

\author[d,e]{Yudai Kazuno}
\affiliation[d]{organization={Graduate School of Comprehensive Human Sciences, University of Tsukuba},
            addressline={1-1-1}, 
            city={Tennodai, Tsukuba},
            postcode={305-8577}, 
            state={Ibaraki},
            country={Japan}}
\affiliation[e]{organization={Urawa Girls' Upper Secondary School},
            addressline={3-8-45}, 
            city={Kishicho, Urawa-ku, Saitama City},
            postcode={335-0064}, 
            state={Saitama},
            country={Japan}}
\author[f]{Masao Mori}
\affiliation[f]{organization={Center for Computational Sciences, University of Tsukuba},
            addressline={1-1-1}, 
            city={Tennodai, Tsukuba},
            postcode={305-8577}, 
            state={Ibaraki},
            country={Japan}}

\begin{abstract}
In the standard cold dark matter (CDM) model, sub-galactic structures hierarchically collide and merge to build up larger structures.
Mergers and collisions between dwarf galaxies and dark matter subhaloes (DMSHs) play an important role in the evolution and formation of structures within a massive galaxy. 
We investigate the collision frequency between DMSHs associated with a massive host galaxy such as the Milky Way.
We analytically estimate the density distribution of DMSH pairs for the relative distance and relative velocity ($r_\mathrm{rel}$-$v_\mathrm{rel}$) and the distance from the centre of the host halo and relative velocity ($r$-$v_\mathrm{rel}$) planes, based on the distribution function of the host halo in the phase space.
Then, we evaluate the collision frequencies of DMSHs by integrating the orbital evolution of DMSHs in Milky-Way-like host haloes selected from cosmological $N$-body simulations. 
The frequency of violent encounters, in which the relative distance of DMSHs is shorter than the sum of scale radii, is averaged as $2.1\times 10^2\,\mathrm{Gyr}^{-1}$. 
Since the time scale of violent encounters, $4.7\,\mathrm{Myr}$, is shorter than the dynamical time of the host halo, collisions between DMSHs occur frequently within the host halo.
Although interactions between DMSHs produce pairs with higher relative velocities, the density distributions of all and colliding pairs between DMSHs provided by numerical results are approximately similar to those of the analytical model neglecting the interactions of DMSHs on $r_\mathrm{rel}$-$v_\mathrm{rel}$ plane for all pairs and $r$-$v_\mathrm{rel}$ plane for colliding pairs. We compare our results with observed colliding dwarf galaxies and provide insight into the abundance of DMSHs.
\end{abstract}




\end{frontmatter}




\section{Introduction}
\label{introduction}
In the standard scenario of galaxy formation based on the $\Lambda$ cold dark matter (CDM) model, collisions and mergers of small-scale structures play a significant role in the formation of larger structures. The CDM model can support the statistical properties of galactic-scale distribution in our Universe.

However, some observational results of the dwarf galaxies are inconsistent with the properties predicted by the CDM model \cite[recent reviews:][] {BullockBoylan-Kolchin_2017_SmallScaleChallengesLCDM_AnnualReviewofAstronomyandAstrophysics, deMartinoEtAl_2020_DarkMattersScale_Universe, SalesEtAl_2022_BaryonicSolutionsChallenges_NatureAstronomy}. 
The missing satellite problem was pointed out as a discrepancy between theoretical predictions and observational results for the number of small-scale systems in the Milky Way (MW)-like host halo with the virial mass $\sim10^{12}\,\mathrm{M_\odot}$ \citep{MooreEtAl_1999_DarkMatterSubstructure_TheAstrophysicalJournalLetters, KlypinEtAl_1999_WhereAreMissing_TheAstrophysicalJournal}.
Numerical simulations based on the $\Lambda$ CDM paradigm show that MW-like haloes have $\gtrsim 100$ dark matter subhaloes (DMSHs) with mass $>10^8\,\mathrm{M_\odot}$, which indicates that the small-scale systems are more abundant than the observed satellites in the MW \citep{IshiyamaEtAl_2008_EnvironmentalEffectSubhalo_PublicationsoftheAstronomicalSocietyofJapan}.
Until the 21st century, only about a tenth of the predicted satellite galaxies had been identified in the MW, 
but improvements in observational techniques have enabled the discovery of many faint dwarf galaxies \citep{McConnachie_2012_OBSERVEDPROPERTIESDWARF_TheAstronomicalJournal, KoposovEtAl_2015_BEASTSSOUTHERNWILD_TheAstrophysicalJournal, BechtolEtAl_2015_EIGHTNEWMILKY_TheAstrophysicalJournal, MartinEtAl_2015_HYDRAIIFAINT_TheAstrophysicalJournalLetters, KimJerjen_2015_HOROLOGIUMIISECOND_TheAstrophysicalJournalLetters, KimEtAl_2015_HEROSDARKHORSE_TheAstrophysicalJournalLetters, LaevensEtAl_2015_SAGITTARIUSIIDRACO_TheAstrophysicalJournal, KirbyEtAl_2015_TRIANGULUMIIPOSSIBLY_TheAstrophysicalJournalLetters, Drlica-WagnerEtAl_2015_EIGHTULTRAFAINTGALAXY_TheAstrophysicalJournal, HommaEtAl_2016_NEWMILKYWAY_TheAstrophysicalJournal, TorrealbaEtAl_2016_SurveyLimitsDiscovery_MonthlyNoticesoftheRoyalAstronomicalSociety, HommaEtAl_2018_SearchesNewMilky_PublicationsoftheAstronomicalSocietyofJapan, KoposovEtAl_2018_SnakeCloudsNew_MonthlyNoticesoftheRoyalAstronomicalSociety, TorrealbaEtAl_2018_DiscoveryTwoNeighbouring_MonthlyNoticesoftheRoyalAstronomicalSociety, MunozEtAl_2018_MegaCamSurveyOuter_TheAstrophysicalJournal, TorrealbaEtAl_2019_HiddenGiantDiscovery_MonthlyNoticesoftheRoyalAstronomicalSociety, HommaEtAl_2019_BootesIVNew_PublicationsoftheAstronomicalSocietyofJapan, MauEtAl_2020_TwoUltrafaintMilky_TheAstrophysicalJournal, CernyEtAl_2023_PegasusIVDiscovery_TheAstrophysicalJournal, SmithEtAl_2024_DiscoveryFaintestKnown_TheAstrophysicalJournal, HommaEtAl_2024_FinalResultsSearch_PublicationsoftheAstronomicalSocietyofJapan}. In the MW, at least 50 satellite dwarf galaxies have been discovered with a range of stellar mass from $\sim10^2\,\mathrm{M_\odot}$ to $\sim10^{9}\,\mathrm{M_\odot}$.

Within the CDM paradigm, several studies have reported on the properties of less massive DMSHs using results from high-resolution cosmological $N$-body simulations
\citep[e.g., ][]{IshiyamaEtAl_2021_UchuuSimulationsData_MonthlyNoticesoftheRoyalAstronomicalSociety, MolineEtAl_2023_LCDMHaloSubstructure_MonthlyNoticesoftheRoyalAstronomicalSociety, KanedaEtAl_2024_UniversalScalingRelation_PublicationsoftheAstronomicalSocietyofJapan, KazunoEtAl_2024_CosmologicalEvolutionDark_PublicationsoftheAstronomicalSocietyofJapan}. 
To explain the abundance discrepancy of satellite dwarf galaxies, the effects of baryonic physical processes
and alternative dark matter models are suggested.
The baryonic processes such as supernova feedback and cosmic reionisation reduce the number of dwarf galaxies by suppressing star formation efficiency. 
The results of cosmological simulations based on the CDM paradigm including baryons predict not only the observed scaling relations of galaxies but also the number of the dwarf galaxies in the MW-like galaxies \citep[e.g. ][]{SawalaEtAl_2016_APOSTLESimulationsSolutions_MonthlyNoticesoftheRoyalAstronomicalSociety, GrandEtAl_2021_DeterminingFullSatellite_MonthlyNoticesoftheRoyalAstronomicalSociety}.
However, \citet{HommaEtAl_2024_FinalResultsSearch_PublicationsoftheAstronomicalSocietyofJapan} mention that the recent discovery of many satellite galaxies in the MW exceeds the predicted number of dwarf galaxies in hydrodynamic cosmological simulations.

By observing dwarf galaxies associated with other galaxies, several studies attempt to provide clues into this problem. 
In the nearest massive galaxy, Andromeda galaxy (M31), $\sim 40$ satellite dwarf galaxies have been identified \citep[][and references therein]{McConnachieEtAl_2018_LargescaleStructureHalo_TheAstrophysicalJournal}, and the latest discovery of a satellite galaxy Pegasus V is reported by \citet{CollinsEtAl_2022_PegasusAndromedaXXXIV_MonthlyNoticesoftheRoyalAstronomicalSociety:Letters}. 
For the third-largest member of the Local Group, M33, which is known as a companion of M31, only three satellite galaxies have been discovered \citep{MartinEtAl_2009_PAndASCUBSDISCOVERY_TheAstrophysicalJournal, Martinez-DelgadoEtAl_2022_PiscesVIIDiscovery_MonthlyNoticesoftheRoyalAstronomicalSociety, OgamiEtAl_2024_TriangulumIVPossible_}.
The number of satellite galaxies in MW-like host galaxies outside the Local Group is also similar to MW \citep{TanakaEtAl_2018_MissingSatelliteProblem_TheAstrophysicalJournal, NashimotoEtAl_2022_MissingSatelliteProblem_TheAstrophysicalJournal}.

Under these situations, it is important for the missing satellite problem to detect the signatures of extremely faint galaxies and DMSHs with little or no stellar component in the CDM paradigm.
In 2015, ultra-diffuse galaxies (UDGs) have been discovered which have extremely low central surface brightness $\mu(g,0)>24\,\mathrm{mag\,arcsec^{-2}}$ and large effective radius $r_\mathrm{eff}>1.5\,\mathrm{kpc}$ \citep{DokkumEtAl_2015_FORTYSEVENMILKYWAYSIZED_TheAstrophysicalJournal, KodaEtAl_2015_APPROXIMAtelYTHOUSANDULTRADIFFUSE_TheAstrophysicalJournal}.
Dragonfly 44, one of the UDGs in the Coma cluster, with a stellar mass of approximately $3 \times 10^8\,\mathrm{M_\odot}$ and a total halo mass of about $\sim10^{12}\,\mathrm{M_\odot}$, exemplifies a dwarf galaxy with suppressed star formation  \citep{DokkumEtAl_2016_HIGHStelLARVELOCITY_TheAstrophysicalJournal}.
Most of the UDGs have a high dark matter fraction, but recent observations report UDGs with extremely low dark matter mass \citep{vanDokkumEtAl_2018_GalaxyLackingDark_Nature, DokkumEtAl_2019_SecondGalaxyMissing_TheAstrophysicalJournal}.

Furthermore, signatures of interactions of small-scale objects are also clues to the missing satellite problem.
In the Andromeda galaxy, \citet{KomiyamaEtAl_2018_StellarStreamHalo_TheAstrophysicalJournal} reported that a stream gap in the North-Western stream is formed by tidal interaction between the Andromeda galaxy and a satellite dwarf galaxy.
The gap structure is predicted as the signature of the DMSH colliding with the stellar stream by using numerical simulations \citep{CarlbergEtAl_2011_DENSITYVARIATIONSNW_TheAstrophysicalJournal, Carlberg_2012_DARKMATTERSUBHALO_TheAstrophysicalJournal}.
By identifying the signatures of interaction between the host halo and DMSHs, it is possible to estimate the collision frequency and indirectly provide insights into the starless DMSHs.

In recent years, the development of observational instruments has allowed observations of more faint structures in the Universe such as the interaction signatures between dwarf galaxies. 
Signatures of mergers and collisions between dwarf galaxies have been detected in the local universe
\cite[e.g.,][]{RichEtAl_2012_TidallyDistortedDwarf_Nature, PaudelEtAl_2015_CASESTUDYTIDAL_TheAstronomicalJournal, AnnibaliEtAl_2016_DDO68FLEA_TheAstrophysicalJournalLetters, PearsonEtAl_2016_LocalVolumeTiNy_MonthlyNoticesoftheRoyalAstronomicalSociety, PaudelSengupta_2017_UGC4703Interacting_TheAstrophysicalJournalLetters, PaudelEtAl_2017_NEXTGENERATIONVIRGO_TheAstrophysicalJournal, PaudelEtAl_2018_CatalogMergingDwarf_TheAstrophysicalJournalSupplementSeries, Kado-FongEtAl_2020_StarFormationIsolated_TheAstronomicalJournal, ChhatkuliEtAl_2023_FormingBlueCompact_MonthlyNoticesoftheRoyalAstronomicalSociety, PaudelEtAl_2023_EarlytypeDwarfGalaxies_MonthlyNoticesoftheRoyalAstronomicalSociety:Letters}.
\citet{PaudelSengupta_2017_UGC4703Interacting_TheAstrophysicalJournalLetters} collected a catalogue including 177 merging dwarf galaxies and categorised the features of systems with shells, stellar streams, loop or antennae structures.
\citet{ChhatkuliEtAl_2023_FormingBlueCompact_MonthlyNoticesoftheRoyalAstronomicalSociety} reported that blue compact dwarf galaxies, characterized by tidal features, are formed through dwarf-dwarf mergers, as evidenced by fitting two-component stellar profiles. Continued improvements in observational techniques are expected to enhance studies of the structural properties of merging galaxies.
On the theoretical side of the collisions between dwarf galaxies, it is initially studied as the formation scenario of the dark-matter-deficient galaxies \citep{Silk_2019_UltradiffuseGalaxiesDark_MonthlyNoticesoftheRoyalAstronomicalSociety:Letters, ShinEtAl_2020_DarkMatterDeficient_TheAstrophysicalJournal, LeeEtAl_2021_DarkMatterDeficient_TheAstrophysicalJournalLetters, OtakiMori_2023_FrequencyDarkMatter_MonthlyNoticesoftheRoyalAstronomicalSociety, LeeEtAl_2024_MultipleBeadsString_TheAstrophysicalJournal}.
\citet{OtakiMori_2023_FrequencyDarkMatter_MonthlyNoticesoftheRoyalAstronomicalSociety} developed a bifurcation sequence of dwarf galaxy formations with analytical methods and numerical simulations of the DMSH collisions. 
Collision events of DMSHs increase the star formation rate of the system and form dwarf galaxies with different dark matter fractions depending on the relative velocity.
Therefore, the frequency of such events is an important indication of the abundance of DMSHs.

In this study, we focus on the collision frequencies between DMSHs and investigate the properties of DMSH pairs in MW-like host haloes using analytical and numerical models.
The paper is structured as follows.
Section \ref{sec: anal} analyses the collision frequencies for the DMSHs and the density distribution of DMSH pairs using the distribution function of a host halo with a Navarro--Frenk--White (NFW) density profile \citep{NavarroEtAl_1996_StructureColdDark_TheAstrophysicalJournal, NavarroEtAl_1997_UniversalDensityProfile_TheAstrophysicalJournal}.
Then we simulate the orbital integration of DMSHs in Milky-Way-like host haloes selected from the cosmological $N$-body simulation, Phi-4096 simulation \citep{IshiyamaEtAl_2021_UchuuSimulationsData_MonthlyNoticesoftheRoyalAstronomicalSociety} and present the properties of DMSH pairs compared to our analytical model in Section \ref{sec: num}.
Finally, in Section \ref{sec: sumdis}, we summarise the conclusion of this paper and discuss the implications of the missing satellite problem by comparing our results with observations.
We adopt a flat $\Lambda$CDM cosmology with $\Omega_0=0.315,\,\Omega_\mathrm{b}=0.049,\,\Omega_\Lambda = 0.685,\,h=0.674,\,n_\mathrm{s}=0.965$, and $\sigma_8=0.811$ in the Planck Collaboration (\citeyear{AghanimEtAl_2020_Planck2018Results_Astronomy&Astrophysics}).

\section{Analytical model} \label{sec: anal}

We estimate the properties of collisions between DMSHs moving within the virial radius of the host halo under dynamical equilibrium, assuming that the velocities of DMSHs follow the velocity distribution function of the host halo. The total energy of a DMSH is expressed as
\begin{gather}
    E=\frac12 v^2+\Phi_\mathrm{NFW}(r),
\end{gather}
moving with velocity $v$ at position $r$ in the gravitational potential $\Phi_\mathrm{NFW}$ of the NFW profile generated by a host halo with mass $M_\mathrm{200,\, host}$. The NFW potential is given by
\begin{gather}
    \Phi_\mathrm{NFW}(r) = -\frac{GM_\mathrm{200,\,host} g(c_\mathrm{host})}{r_\mathrm{s,\,host}}\frac{\ln{(1+x)}}{x},\label{eq:NFWpot}\\
    x =\frac{r}{r_\mathrm{s,\,host}},\\
    g(x)=\frac{1}{\ln(1+x)-x/(1+x)}.
\end{gather}
where $G$ is the gravitational constant, $c_\mathrm{host}=R_\mathrm{200,\,host}/r_\mathrm{s,\,host}$ is the concentration of a host halo, $R_\mathrm{200,\,host}=\left({3M_\mathrm{200,\,host}}/{4\pi \rho_{200}}\right)^{1/3}$ is a virial radius of a host halo, $r_\mathrm{s,\,host}$ is a scale radius of the host halo, and $\rho_{200}$ is $200$ times the critical density of the universe.
The studies demonstrate that the concentrations tightly correlate with the mass such as the $c\text{--}M$ relation \citep[e.g., ][]{BullockEtAl_2001_ProfilesDarkHaloes_MonthlyNoticesoftheRoyalAstronomicalSociety, PradaEtAl_2012_HaloConcentrationsStandard_MonthlyNoticesoftheRoyalAstronomicalSociety, DiemerJoyce_2019_AccuratePhysicalModel_TheAstrophysicalJournal, IshiyamaAndo_2020_AbundanceStructureSubhaloes_MonthlyNoticesoftheRoyalAstronomicalSociety, IshiyamaEtAl_2021_UchuuSimulationsData_MonthlyNoticesoftheRoyalAstronomicalSociety, KanedaEtAl_2024_UniversalScalingRelation_PublicationsoftheAstronomicalSocietyofJapan}.
We adopt the $c\text{-}M$ relation proposed by \citet{DiemerJoyce_2019_AccuratePhysicalModel_TheAstrophysicalJournal},
\begin{gather}
    c=C\left(\alpha_{\mathrm{eff}}\right) \times \tilde{G}\left(\frac{A\left(n_{\mathrm{eff}}\right)}{\nu}\left[1+\frac{\nu^2}{B\left(n_{\mathrm{eff}}\right)}\right]\right) \label{eq:c-M},
\end{gather}
where the parameters and the functions are described in \ref{app:c-M}.
In this paper, the fitting parameters of equation \eqref{eq:c-M} are set as 
$\kappa=1.10,\,a_0=2.30,\,a_1=1.64,\,b_0=1.72,\,b_1=3.60,$ and $c_\alpha=0.32$, which are fitted for all haloes using $V_\mathrm{max}$ method in the Uchuu simulations \citep{IshiyamaEtAl_2021_UchuuSimulationsData_MonthlyNoticesoftheRoyalAstronomicalSociety} {to compare the numerical results described in Section \ref{sec: num}}.
The concentration of a MW-like host halo with the mass of $M_\mathrm{200}=10^{12}\,\mathrm{M_\odot}$ corresponds to $c_\mathrm{host}=8.14$, which is calculated using the \texttt{COLOSSUS} package developed by \citet{Diemer_2018_COLOSSUSPythonToolkit_TheAstrophysicalJournalSupplementSeries}.
In \ref{app:anal}, we show a result of the analytical model for $c_\mathrm{host}=12.0$ and discuss the dependency of the concentration parameter.

Since we consider DMSHs bound to the host galaxy, the energy of DMSHs $E$ is always negative.
Here, the energy and the potential are expressed as $\mathcal{E}=-E\text{ and }\Psi=-\Phi_\mathrm{NFW}$, respectively.
The distribution function given a spherical density profile can be calculated by Eddington's formula \citep{BinneyTremaine_2008_GalacticDynamicsSecond_GalacticDynamics:SecondEditionbyJamesBinneyandScottTremaine.ISBN978-0-691-13026-2HB.PublishedbyPrincetonUniversityPressPrincetonNJUSA2008.},
\begin{gather}
    f(\mathcal{E})=\frac{1}{\sqrt{8}\pi^2}\left[\frac{1}{\sqrt{\mathcal{E}}}\left(\frac{\mathrm{d}\nu}{\mathrm{d}\Psi}\right)_{\Psi=0}+\int^{\mathcal{E}}_0\frac{\mathrm{d}^2\nu}{\mathrm{d}\Psi^2}\frac{\mathrm{d}\Psi}{\sqrt{\mathcal{E}-\Psi}}\right],
\end{gather}
where $\nu$ is the probability density. For an NFW profile, it is written by
\begin{gather}
    \nu(r)=\frac{\rho_\mathrm{NFW}(r)}{M_\mathrm{200,\,host}}=\frac{g(c_\mathrm{host})}{4\pi r_\mathrm{s,\,host}^3}\frac{1}{x(1+x)^2},
\end{gather}
where $\rho_\mathrm{NFW}$ is the density distribution of the NFW profile expressed by
\begin{gather}
    \rho_\mathrm{NFW}(r) = \frac{c^3g(c)}{3x(1+x)} \rho_{200}.
\end{gather}

In analytical models, we assumed that the velocity distribution of DMSHs follows the distribution function $f(\mathcal{E})$ under the host halo to simplify the motion of DMSHs.
We note that it is not trivial whether the system of DMSHs satisfies this condition and needs to be verified in future work.

\subsection{Relative distance distribution between dark matter subhaloes}
We calculate the probability distribution for the relative distances and velocities between two DMSHs positioned at $\bm{r}_1$ and $\bm{r}_2$, moving at velocities $\bm{v}_1$ and $\bm{v}_2$, respectively. In this calculation, interactions between the two DMSHs are ignored. Accordingly, the distribution function for two DMSHs can be expressed as $f(\mathcal{E}_1) f(\mathcal{E}_2) \mathrm{d}^3 \bm{r}_1 \mathrm{d}^3 \bm{v}_1 \mathrm{d}^3 \bm{r}_2 \mathrm{d}^3 \bm{v}_2$.

The position and velocity corresponding to the centre of mass and the relative coordinates are indicated by $\bm{r}_\mathrm{cm}=(\bm{r}_1+\bm{r}_2)/2$, $\bm{v}_\mathrm{cm}=(\bm{v}_1+\bm{v}_2)/2$, $\bm{r}_\mathrm{rel}=\bm{r}_1-\bm{r}_2$ and $\bm{v}_\mathrm{rel}=\bm{v}_1-\bm{v}_2$, respectively.
By integrating the distribution function in the phase space of the centre-of-mass coordinate, the probability function of relative distances and relative velocities for an isotropic distribution is defined as
\begin{align}
    P_\mathrm{rel}(r_\mathrm{rel},v_\mathrm{ref})=&\int_0^\infty \mathrm{d}r_\mathrm{cm}r_\mathrm{cm}^2\int_{-1}^1\mathrm{d}y\int_0^\infty \mathrm{d}v_\mathrm{cm}v_\mathrm{cm}^2\int_{-1}^1\mathrm{d}z \nonumber \label{eq:Prel}\\
    &\times 64\pi^4 r_\mathrm{rel}^2v_\mathrm{rel}^2  \,f(\mathcal{E}_1) f(\mathcal{E}_2),
\end{align}
\begin{align}
    \mathcal{E}_1 &= \Psi_\mathrm{NFW}\left(|\bm{r}_1|\right)-\frac12 |\bm{v}_1|^2\nonumber\\
    &=\Psi_\mathrm{NFW}\left(\sqrt{r_\mathrm{cm}^{2} + r_\mathrm{cm} r_\mathrm{rel} y+{r_\mathrm{rel}^{2}}/{4}}\right) \nonumber\\
    &\qquad -\frac12(v_\mathrm{cm}^{2} + v_\mathrm{cm}v_\mathrm{rel} z+{v_\mathrm{rel}^{2}}/{4}),\\
    \mathcal{E}_2 &= \Psi_\mathrm{NFW}(|\bm{r}_2|)-\frac12 |\bm{v}_2|^2\nonumber\\
    &= \Psi_\mathrm{NFW}\left(\sqrt{r_\mathrm{cm}^{2} - r_\mathrm{cm} r_\mathrm{rel} y+{r_\mathrm{rel}^{2}}/{4}}\right)\nonumber \\
    & \qquad -\frac12(v_\mathrm{cm}^{2} - v_\mathrm{cm}v_\mathrm{rel} z+{v_\mathrm{rel}^{2}}/{4}),
    \end{align}
where $r_\mathrm{cm}=|\bm{r}_\mathrm{cm}|,\,r_\mathrm{rel}=|\bm{r}_\mathrm{rel}|,\, \bm{r}_\mathrm{cm}\cdot\bm{r}_\mathrm{rel}=r_\mathrm{cm}r_\mathrm{rel}y$, $y=\cos\theta_r$ $v_\mathrm{cm}=|\bm{v}_\mathrm{cm}|,\,v_\mathrm{rel}=|\bm{v}_\mathrm{rel}|,\,\bm{v}_\mathrm{cm}\cdot\bm{v}_\mathrm{rel}=v_\mathrm{cm}v_\mathrm{rel}z$, and $z = \cos\theta_v$. We define $\theta_r$ between $\bm{r}_\mathrm{cm}$ and $\bm{r}_\mathrm{rel}$, and $\theta_v$ between $\bm{v}_\mathrm{cm}$ and $\bm{v}_\mathrm{rel}$.
The probability distribution is normalized such that
$\int P_\mathrm{rel}(r_\mathrm{rel},v_\mathrm{rel}),\mathrm{d} r_\mathrm{rel},\mathrm{d} v_\mathrm{rel} = 1$.

\begin{figure}
    \centering
    \includegraphics[width=\columnwidth]{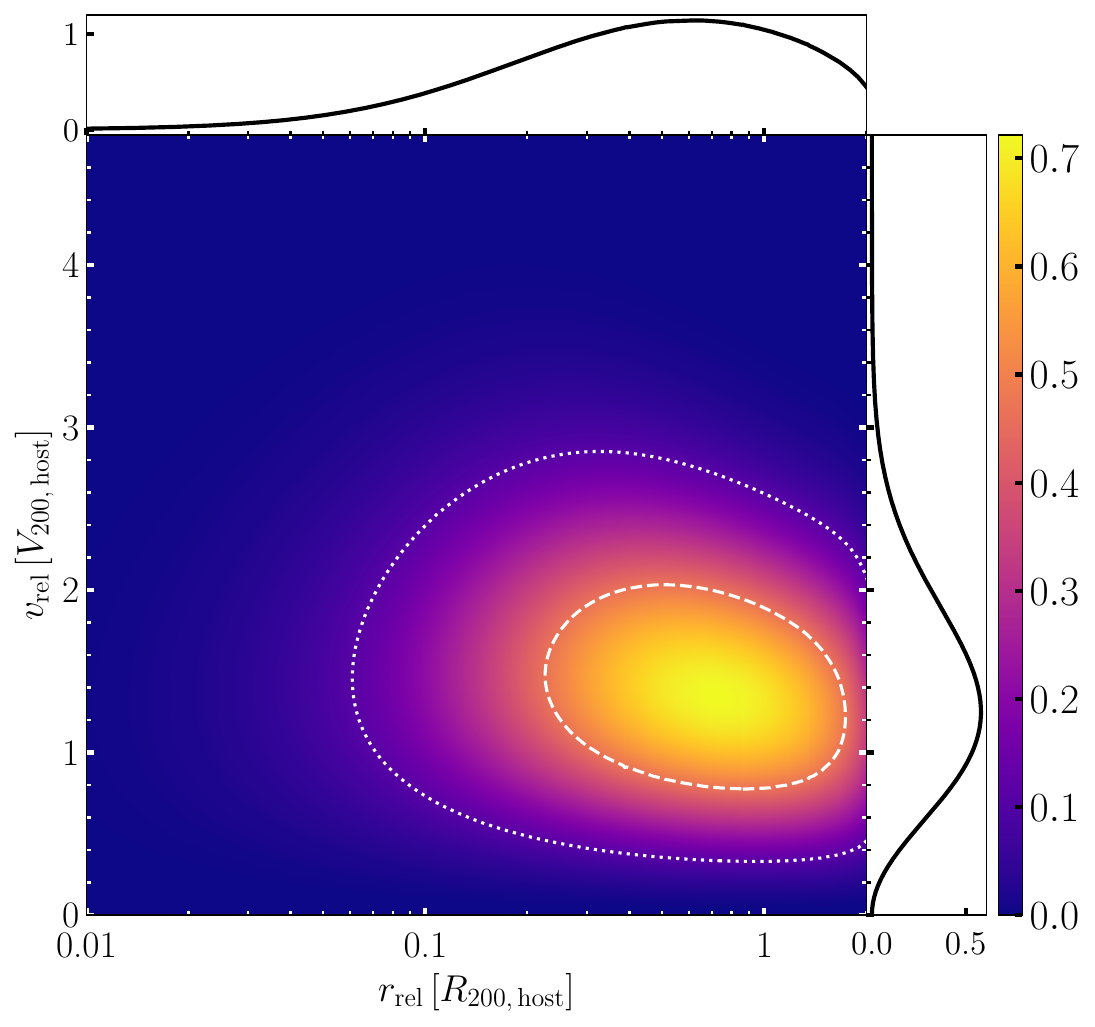}
    \caption{Probability density $P_\mathrm{rel}(r_\mathrm{rel},v_\mathrm{ref})$ for the relative distance $r_\mathrm{rel}$ and relative velocity $v_\mathrm{rel}$ in the host halo with $c_\mathrm{host}=8.14$ calculated by equation \eqref{eq:Prel}. The white dashed and dotted contours correspond to the regions that contain 39\% ($1\sigma$) and 86\% ($2\sigma$) of the total probability mass from the highest value probability density, respectively. The top sub-panel indicates the probability depending on the relative distance, which is integrated by the relative velocity. The right sub-panel indicates the probability depending on the relative velocity, which is integrated by the relative distance.}
    \label{fig: Prel}
\end{figure}

We use the Monte Carlo integration method to obtain the result of equation \eqref{eq:Prel}.
Fig. \ref{fig: Prel} shows the probability distribution for the relative distance and relative velocity of two DMSHs in the host halo with $c_\mathrm{host}=8.14$. 
The horizontal axis corresponds to relative distance $r_\mathrm{rel}$ normalised by the virial radius $R_{200,\,\mathrm{host}}$ and the vertical axis corresponds to the relative velocity $v_\mathrm{rel}$ normalised by the circular velocity $V_{200,\,\mathrm{host}}=\sqrt{GM_{200,\,\mathrm{host}}/R_{200,\,\mathrm{host}}}$. 
The white dashed and dotted contours represent regions enclosing 39\% ($1\sigma$) and 86\% ($2\sigma$) of the total probability mass, calculated from the highest density points, respectively.

The probability of two DMSHs is the highest for a relative distance of $0.619\,R_{200,\,\mathrm{host}}$ and a relative velocity of $1.25\,V_{200,\,\mathrm{host}}$.
The probability density distribution extends elliptically from the position of the peak value. 
It also indicates that the probability of the DMSH pairs with the relative distances of $\lesssim0.1\,R_{200,\,\mathrm{host}}$ and the relative velocities of $\gtrsim 3\,V_{200,\,\mathrm{host}}$ or less is relatively low. 

For a MW-like host halo with $M_{200,\,\mathrm{host}}=10^{12}\,\mathrm{M_\odot}$, the virial radius and circular velocity at this radius are $R_{200,\,\mathrm{host}} = 212 \,\mathrm{kpc}$ and $V_{200,\,\mathrm{host}} = 143\,\mathrm{km\,s^{-1}}$, respectively.
Therefore, within the phase-space distribution, the DMSH pairs with the relative distance of $131\,\mathrm{kpc}$ and relative velocity of $179 \,\mathrm{km\,s^{-1}}$ have the highest probability. 
However, such pairs cannot be determined to collide because the relative distances are large compared to the typical scale radii of the DMSHs.

\subsection{Collision frequency in a host halo}

\begin{figure}
    \centering
    \includegraphics[width=\columnwidth]{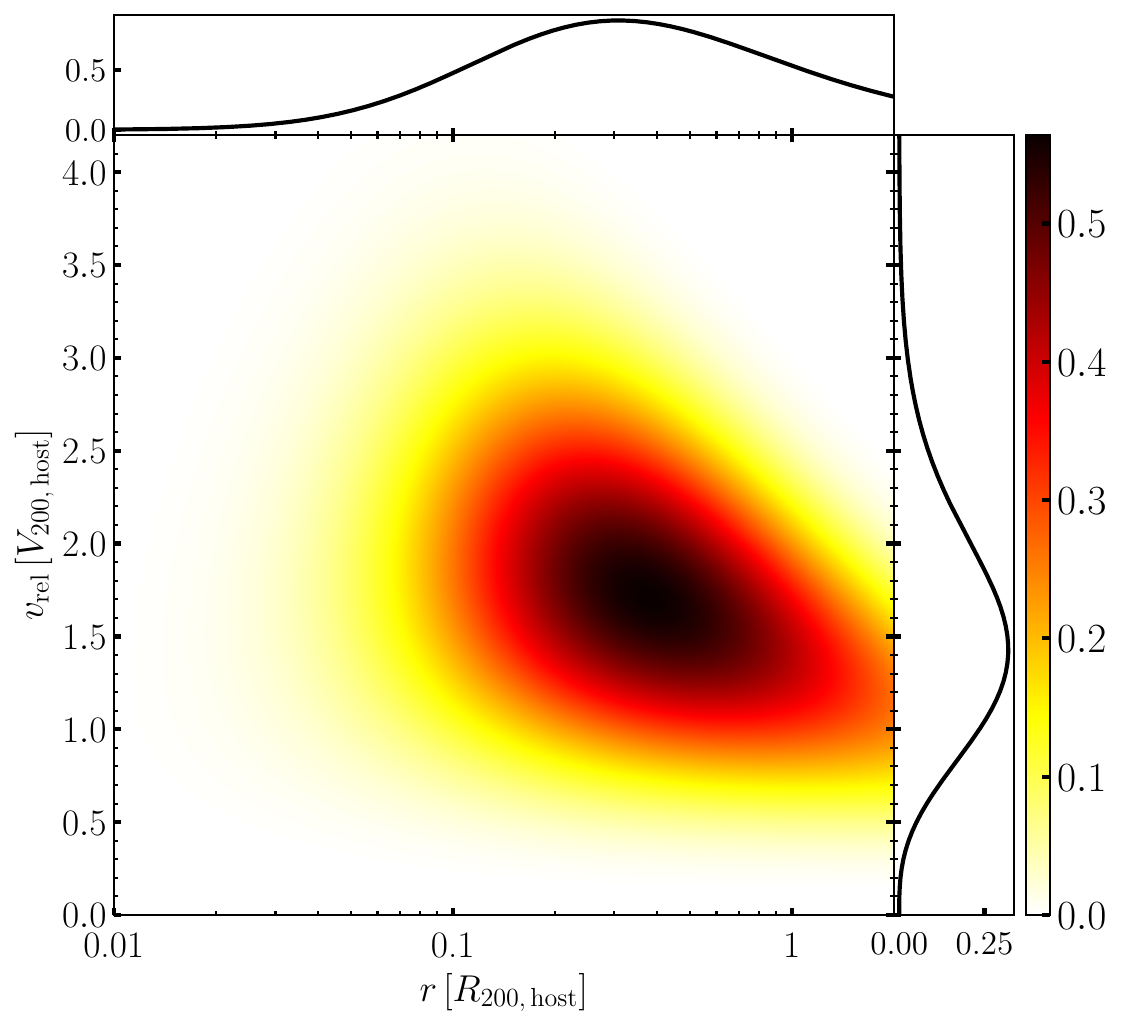}
    \caption{Density distribution of collision frequency ${c_\mathrm{sub}^2\mathrm{d}k_\gamma}/({N_\mathrm{sub}^2\eta^2\mathrm{d} r\,\mathrm{d}v_\mathrm{rel}\,\mathrm{d}t})$ between DMSHs within a host galaxy for $\gamma = 1.5$ and $c_\mathrm{host}=8.14$. The top sub-panel indicates the dependence of the collision frequency on the radius of a host halo. The right sub-panel indicates the dependence of the collision frequency on the relative velocity.}
    \label{fig: col_freq}
\end{figure}

Next, we evaluate the collision frequency between two DMSHs within a host halo by applying the number density profile of the DMSHs introduced by \citet{HanEtAl_2016_UnifiedModelSpatial_MonthlyNoticesoftheRoyalAstronomicalSociety}.
In the analytical model of the collision frequency, \citet{OtakiMori_2023_FrequencyDarkMatter_MonthlyNoticesoftheRoyalAstronomicalSociety} define
the expected value of the relative velocity between two DMSHs as
\begin{gather}
    E[v_{r,\,\mathrm{rel}}] = \int \mathrm{d}v_{r,\,\mathrm{rel}}, \quad  \mathrm{d}v_{r,\,\mathrm{rel}} \equiv v_\mathrm{rel} P_{r,\,\mathrm{rel}}(v_\mathrm{rel})\mathrm{d}v_\mathrm{rel}.
\end{gather}
Here, $P_{r,\,\mathrm{rel}}(v_\mathrm{rel})$ is the probability distribution of the relative velocity at a position $r$ in the host halo for isotropic velocities, which is expressed as
\begin{gather}
    P_{r,\mathrm{rel}}(v_\mathrm{rel}) 
    = \frac{8\pi^2v_\mathrm{rel}^2}{\nu(r)^2}\int_0^\infty\mathrm{d}{v}_\mathrm{cm} v_\mathrm{cm}^2 \int_{-1}^{1}\mathrm{d} z \,f(\mathcal{E}_{r,1})f(\mathcal{E}_{r,2}),\label{eq:Prrel} \\
    \mathcal{E}_{r,1} = \Psi(r) -\frac12(v_\mathrm{cm}^2+v_\mathrm{rel}^2/4+v_\mathrm{cm}v_\mathrm{rel}z),\\
    \mathcal{E}_{r,2} = \Psi(r) -\frac12(v_\mathrm{cm}^2+v_\mathrm{rel}^2/4-v_\mathrm{cm}v_\mathrm{rel}z).
\end{gather}
For the number density of DMSHs $n(r)$ , the number of collisions in a volume $4\pi r^2 \mathrm{d}r$ during a time $\mathrm{d}t$ is derived by
\begin{gather}
    \mathrm{d}k = 4\pi r^2n(r)^2\frac\sigma2\,\mathrm{d}r\,\mathrm{d}v_{r,\,\mathrm{rel}}\,\mathrm{d}t,
\end{gather}
where $\sigma$ is the cross-section of colliding DMSHs. We set $\sigma = \pi r_\mathrm{s,\, sub}^2$ for a scale radius of the DMSH.

\citet{OtakiMori_2023_FrequencyDarkMatter_MonthlyNoticesoftheRoyalAstronomicalSociety} simply assumed that the spatial distribution of the  DMSHs follows the NFW function. On the other hand, as an extended model incorporating the effect of tidal stripping onto the DMSHs suffered from the host halo, \citet{HanEtAl_2016_UnifiedModelSpatial_MonthlyNoticesoftheRoyalAstronomicalSociety} proposed the following function,

\begin{gather}
    \frac{\mathrm{d}N(M_\mathrm{sub}, \,r)}{\mathrm{d}\ln m\,\mathrm{d}V}\propto M_\mathrm{sub}^{-\alpha}r^\gamma \rho_\mathrm{host}(r), \label{eq:han}
\end{gather}
where $M_\mathrm{sub}$ is the DMSH mass, $\rho_\mathrm{host}(r)$ is the density profile of the host halo, and the parameter $\gamma$ characterizes the spatial distribution of DMSHs.

It is assumed that the infall mass function of DMSHs is expressed as a power law function, the spatial distribution of DMSHs traces the density profile of the host halo, and the mass of the DMSH is tidally stripped in proportion to the power of the distance from the host centre.
In equation \eqref{eq:han}, $\gamma=0$ corresponds to the model adopted by \citet{OtakiMori_2023_FrequencyDarkMatter_MonthlyNoticesoftheRoyalAstronomicalSociety}. 
Here we adopt $\gamma=1.5$ averaged over the fitted parameter for the number profile of DMSHs within the selected MW-like host halo in the {Phi-4096} simulation (see \S \ref{sec: setup} and Table \ref{tab: Phi4096_prop}). {In \ref{app:anal}, we show a result of the model for $\gamma=1$ and discuss the dependency of $\gamma$ on our model.}

In this analytical estimation, we assume the density distribution of the host halo follows an NFW profile. 
Accordingly, we redefine the number density profile of DMSHs as
\begin{align}
    n_\gamma(r) &= \frac{N_\mathrm{sub} r^\gamma \nu_\mathrm{host}(r)}{A_\gamma r_\mathrm{s,\,\mathrm{host}}^\gamma g(c_\mathrm{host})}, \nonumber\\
    &= \frac{N_\mathrm{sub}}{4\pi r_{\mathrm{s,\,host}}^3A_\gamma(c_\mathrm{host})}\frac{x^\gamma}{x(1+x)^2} \label{eq:number_density},
\end{align}
where $N_\mathrm{sub}$ is the total number of DMSHs in the host halo and $x = r/r_\mathrm{s,\, host}$. $A_\gamma$ is a normalization given by
\begin{gather}
    A_\gamma(c_\mathrm{host}) = \int_0^{c_\mathrm{host}}\frac{x^{\gamma + 1}}{(1+x)^2}\mathrm{d}x,
\end{gather}

The number of collisions, accounting for tidal effects, is calculated as
\begin{align}
    \mathrm{d}{k_\gamma} &=  4\pi r^2n_\gamma^2(r)\frac\sigma2v_\mathrm{rel}P_{r,\,\mathrm{rel}}\,\mathrm{d}r\,\mathrm{d}v_\mathrm{rel}\,\mathrm{d}t,\nonumber\\
    &=\frac{N_\mathrm{sub}^2\eta^2}{8R_{200,\,\mathrm{host}}^2c_\mathrm{sub}^2}\frac{c_\mathrm{host}^4}{A_\gamma^2(c_\mathrm{host})}\frac{x^{2\gamma}}{(1+x)^4}v_\mathrm{rel}P_{r,\,\mathrm{rel}}\mathrm{d}r\,\mathrm{d}v_\mathrm{rel}\,\mathrm{d}t,
\end{align}
where the parameter $\eta$ represents the ratio of the virial radius of the host halo to that of a DMSH,
\begin{gather}
    \eta = \frac{R_\mathrm{200,\,sub}}{R_\mathrm{200,\,host}}=\left(\frac{M_\mathrm{sub}}{M_\mathrm{host}}\right)^{1/3}.
\end{gather}

The distribution of collision frequency, normalized by the parameters of colliding DMSHs, is expressed as
\begin{gather}
    \frac{c_\mathrm{sub}^2}{N_\mathrm{sub}^2\eta^2}\frac{\mathrm{d}k_\gamma}{\mathrm{d}r\,\mathrm{d}v_\mathrm{rel}\,\mathrm{d}t} = 
    \frac{c_\mathrm{host}^4}{8R_{200,\,\mathrm{host}}^2A_\gamma^2(c_\mathrm{host})}\frac{x^{2\gamma}}{(1+x)^4}v_\mathrm{rel}P_{r,\,\mathrm{rel}}.
\end{gather}

Fig. \ref{fig: col_freq} shows the distribution of collision frequency ${c_\mathrm{sub}^2\mathrm{d}k_\gamma}/({N_\mathrm{sub}^2\eta^2\mathrm{d} r\,\mathrm{d}v_\mathrm{rel}\,\mathrm{d}t})$ for $\gamma = 1.5$ and $c_\mathrm{host}=8.14$. The horizontal and vertical axes represent the distance from the centre of the host halo normalised by $R_{200,\,\mathrm{host}}$ and the relative velocity between two DMSHs normalised by $V_{200,\,\mathrm{host}}$, respectively.
The upper and right sub-panels show ${c_\mathrm{sub}^2\mathrm{d}k_\gamma}/({N_\mathrm{sub}^2\eta^2\mathrm{d} r\,\mathrm{d}t})$ and ${c_\mathrm{sub}^2\mathrm{d}k_\gamma}/({N_\mathrm{sub}^2\eta^2\,\mathrm{d}v_\mathrm{rel}\,\mathrm{d}t})$, respectively.
The dispersion of the relative velocity distribution of DMSH collisions becomes large at around $0.2R_{200,\,\mathrm{host}}$. On the other hand, at the outer region $\gtrsim 1R_{200,\,\mathrm{host}}$ of the host halo, the relative velocity dispersion tends to decrease.
Collision events occur most frequently at the position $r_\mathrm{peak}=0.310\,R_{200,\,\mathrm{host}}$ with the relative velocity $v_\mathrm{rel,\,peak}=1.43\,V_{200,\,\mathrm{host}}$.
For the MW-like host halo with $M_{200,\,\mathrm{host}}=10^{12}\,\mathrm{M_\odot}$, the peak position and the peak relative velocity correspond to $r_\mathrm{peak} = 65.6\,\mathrm{kpc}$ and $v_\mathrm{rel,\,peak}=204\,\mathrm{km\,s^{-1}}$, respectively. 
Compared to the previous study \citep{OtakiMori_2023_FrequencyDarkMatter_MonthlyNoticesoftheRoyalAstronomicalSociety}, this model includes the number profile of the DMSHs modelled by \citet{HanEtAl_2016_UnifiedModelSpatial_MonthlyNoticesoftheRoyalAstronomicalSociety}. Therefore, collision frequency within the inner region $(r\lesssim0.1\, R_{200,\,\mathrm{host}})$ rapidly decreases due to the tidal disruption of DMSHs.

The total collision frequency within the host halo is defined as
\begin{equation}
    f_{\mathrm{col},\,\gamma}\equiv \frac{\mathrm{d}k_\gamma}{\mathrm{d}t} = \frac{N_\mathrm{sub}\eta^2}{c_\mathrm{sub}^2}\iint\left(\frac{\mathrm{d}k_\mathrm{\gamma}}{\mathrm{d}r^\prime\,\mathrm{d}v_\mathrm{rel}^\prime\, \mathrm{d}t}\right)\mathrm{d}r^\prime\,\mathrm{d}v_\mathrm{rel}^\prime.
\end{equation}

When DMSHs with a mass of $M_\mathrm{sub}=10^9\,\mathrm{M_\odot}$ move within an MW-like host halo, the collision frequency and time scale are calculated as follows:

\begin{gather}
    f_\mathrm{col,\,\gamma}=3.71\left(\frac{N_\mathrm{sub}}{500}\right)^{2}\,\mathrm{Gyr^{-1}},\\
    \tau_\mathrm{col,\,\gamma}= f_\mathrm{col,\,\gamma}^{-1} = 270 \left(\frac{N_\mathrm{sub}}{500}\right)^{-2}\,\mathrm{Myr},
\end{gather}
with $\gamma=1.5$, respectively. Here, the concentration of colliding DMSH, $c_\mathrm{sub}=14.7$, is given by equation \eqref{eq:c-M}, which corresponds to $10^9\,\mathrm{M_\odot}$.
Since the collision time scale is 
comparable to the typical dynamical time scale of dwarf galaxies ($\sim100\,\mathrm{Myr}$), the result implies that DMSHs with the same mass frequently collide within the host halo.

\section{Numerical method} \label{sec: num}
Our simple analytical models provide the probability distributions of DMSHs by neglecting interactions between DMSHs.
In this section, we perform the orbital integration of DMSHs moving within a MW-like host halo to analyze the positions and velocities of colliding DMSHs. Then, we derive the collision frequency between DMSHs and the probability distribution for relative velocities and relative distances.

\subsection{Simulation set-up} \label{sec: setup}

\begin{table*}
\centering
\caption[Properties of Milky Way-like host halos in the Phi-4096 simulation.]{Properties of Milky Way-like host halos in the Phi-4096 simulation: serial number of the host halo identified in the Phi-4096 simulation, virial mass $M_\mathrm{vir,\, host}$, virial radius $R_\mathrm{vir,\, host}$, scale radius $r_\mathrm{s,\, host}$, concentration $c$, number of dark matter subhaloes $N_\mathrm{sub}$ with virial mass $>10^6\,\mathrm{M_\odot}$ within $2R_\mathrm{vir,\, host}$, and the power exponent $\gamma$ of the subhalo number density profile for a distance from center of the host halo, and integration time $\tau_\mathrm{sim}$ corresponding to the increasing timescale of host halo mass from $0.99M_0$ to $M_0$.}
\label{tab: Phi4096_prop}
\begin{tabular}{cccccccc}
\hline
No. &
  $M_\mathrm{200,\,host}\,\mathrm{[10^{11}\,M_\odot]}$ &
  $R_\mathrm{200,\,host}\,\mathrm{[kpc]}$ &
  $r_\mathrm{s,\,host}\,\mathrm{[kpc]}$ &
  $c$ &
  $N_\mathrm{sub}(>10^6\,\mathrm{M_\odot})$ &
  $\gamma$ &
  $\tau_\mathrm{sim}\,\mathrm{[Myr]}$ \\ \hline
1  & $24.4$ & $283$ & $24.1$ & $11.7$ & $28876$ & $1.82$ & $326$  \\
2  & $16.6$ & $249$ & $23.6$ & $10.5$ & $22480$ & $1.67$ & $320$  \\
3  & $15.4$ & $243$ & $20.8$ & $11.7$ & $16170$ & $1.47$ & $340$  \\
4  & $16.0$ & $246$ & $23.9$ & $10.3$ & $20001$ & $1.72$ & $132$  \\
5  & $14.7$ & $239$ & $26.6$ & $9.01$ & $19762$ & $1.50$ & $296$  \\
6  & $14.6$ & $238$ & $28.5$ & $8.35$ & $18588$ & $1.57$ & $232$  \\
7  & $7.34$ & $190$ & $21.8$ & $8.71$ & $12483$ & $1.59$ & $164$  \\
8  & $13.1$ & $230$ & $35.5$ & $6.47$ & $20212$ & $1.54$ & $157$  \\
9  & $8.11$ & $196$ & $69.8$ & $2.81$ & $17626$ & $1.07$ & $157$  \\
10 & $11.1$ & $218$ & $26.4$ & $8.24$ & $18004$ & $1.45$ & $253$  \\
11 & $10.2$ & $212$ & $21.2$ & $9.99$ & $15694$ & $1.56$ & $275$  \\
12 & $11.1$ & $218$ & $25.4$ & $8.57$ & $16369$ & $1.46$ & $213$  \\
13 & $11.2$ & $218$ & $18.5$ & $11.8$ & $14249$ & $1.64$ & $375$  \\
14 & $11.1$ & $218$ & $28.9$ & $7.53$ & $15482$ & $1.30$ & $218$  \\
15 & $9.49$ & $207$ & $26.0$ & $7.94$ & $14123$ & $1.40$ & $233$  \\
16 & $8.66$ & $200$ & $25.1$ & $8.01$ & $12873$ & $1.22$ & $267$  \\
17 & $7.83$ & $194$ & $28.7$ & $6.75$ & $11932$ & $1.40$ & $183$  \\
18 & $6.61$ & $183$ & $17.5$ & $10.5$ & $7463$  & $1.82$ & $237$  \\
19 & $6.05$ & $178$ & $25.0$ & $7.10$ & $6819$  & $1.39$ & $263$  \\
20 & $6.19$ & $179$ & $25.6$ & $7.00$ & $6877$  & $1.52$ & $85.4$ \\
21 & $5.81$ & $175$ & $23.8$ & $7.36$ & $8016$  & $1.56$ & $203$  \\
22 & $4.97$ & $167$ & $28.8$ & $5.79$ & $8641$  & $1.32$ & $220$  \\
23 & $5.68$ & $174$ & $16.5$ & $10.5$ & $8349$  & $1.66$ & $310$  \\
24 & $5.18$ & $169$ & $18.8$ & $8.98$ & $7282$  & $1.71$ & $242$  \\
25 & $5.18$ & $169$ & $16.2$ & $10.4$ & $7104$  & $1.83$ & $349$  \\
26 & $4.94$ & $166$ & $12.6$ & $13.2$ & $4652$  & $2.00$ & $326$  \\
27 & $4.37$ & $160$ & $20.9$ & $7.63$ & $6078$  & $1.34$ & $218$  \\ \hline
\end{tabular}
\end{table*}

We select 27 MW-like host haloes in the Phi-4096 simulations\footnote{In the original Phi-4096 simulation, the cosmological parameters are adopted as $\Omega_0=0.31,\,\Omega_\mathrm{b}=0.048,\,\Omega_\Lambda = 0.69,\,h=0.68,\,n_\mathrm{s}=0.96,\text{ and } \sigma_8=0.83$, but we adopt the parameters from the Planck Collaboration (see text).} \citep{IshiyamaEtAl_2021_UchuuSimulationsData_MonthlyNoticesoftheRoyalAstronomicalSociety}. The particle resolution of the Phi-4096 simulation is $5.13\times 10^3\,h^{-1}\,\mathrm{M_\odot}$.
The range of virial masses selected as host haloes is 
$5\times 10^{11}\,\mathrm{M_\odot}<M_\mathrm{200,\,host}<3\times10^{12}\,\mathrm{M_\odot}$.
The DMSHs are selected with masses greater than $10^6\,\mathrm{M_\odot}$ and within $2R_\mathrm{200,\,host}$ of each host halo at redshift $z=0$. 
Table \ref{tab: Phi4096_prop} lists properties of MW-like host haloes selected from the Phi-4096 simulation and the number of DMSH with a fitting parameter $\gamma$ parameterized in equation \eqref{eq:number_density} to characterize distribution properties of DMSHs at $z=0$.

As initial conditions, the position of the host halo is set to the origin in each simulation at redshift $z=0$.
We conduct 27 $N$-body simulation models. In these, each DMSH is treated as a particle with a point mass within the NFW potential generated by the host halo, and the simulation is run backwards in time, with a time step $\Delta t \rightarrow -\Delta t$. {The mass, initial positions, and initial velocities of each DMSH particle are obtained from the DMSH catalogue at $z=0$ using \texttt{ROCKSTAR} \citep{BehrooziEtAl_2012_ROCKSTARPHASESPACETEMPORAL_TheAstrophysicalJournal} and \texttt{CONSISTENT TREES} \citep{Behroozi_2013}.}
Since our simulation neglects the changes in the potential field due to the mass evolution of the host halo, orbital integration is performed up to the time when the mass of the host galaxy is 99\% of the mass at $z=0$, as determined from the merger tree.
The mass evolution of the host halo is fitted by the exponential function with one parameter $\alpha$,
\begin{gather}
    M(a) = M_0\, \mathrm{e}^{-\alpha z},
\end{gather}
where scale factor $a = (1+z)^{-1}$ and $M_0$ is the host halo mass $z=0$ \citep{WechslerEtAl_2002_ConcentrationsDarkHalos_TheAstrophysicalJournal}. We defined the orbital integration time $\tau_\mathrm{sim} = t_{100} - t_{99}$, where $t_{100}$ and $t_{99}$ are corresponding to $a(t_{100})=1$ and $M(a(t_{99})) = 0.99 M_0$, respectively.
The integration times of simulation $\tau_\mathrm{sim}$ for each host halo are listed in Table \ref{tab: Phi4096_prop}.

The equation of motion for the $i$-th DMSH is calculated by 
\begin{gather}
    \frac{\mathrm{d}^2\bm{r}_i}{\mathrm{d}t^2} = -\sum_{j\neq i}^{N_\mathrm{sub}} \frac{Gm_j \bm{r}_{ij}}{r_{ij}^3} - \nabla \Phi_\mathrm{NFW, \,host}({r}_i),
\end{gather}
where $\bm{r}_{ij} = \bm{r}_i-\bm{r}_j$, $r_{ij}=|\bm{r}_{ij}|$, $m_j$ is the mass of DMSHs and $N_\mathrm{sub}$ is a number of DMSHs. The gravitational force of a host halo is calculated as a static NFW potential,
\begin{gather}
    \nabla\Phi_\mathrm{NFW,\,host}(r) = \frac{GM_\mathrm{200,\,host} g(c_\mathrm{host})}{g(x)}\frac{\bm{r}}{r^3},\label{eq:NFWforce}
\end{gather}
where $r=|\bm{r}|$ and $x=r/r_\mathrm{s,\,host}$.

The orbital evolution of DMSHs is integrated with $\text{P(EC)}^n$ Hermite integrator introduced by \citet{KokuboEtAl_1998_TimesymmetricHermiteIntegrator_MonthlyNoticesoftheRoyalAstronomicalSociety}. 
This scheme is a fourth-order time-symmetric integrator with $n$ iterations between the evaluation (E) and correction (C) steps after the prediction (P) step to reduce the integration error. 
In this paper, we set the number of iterations as $n=1$, which corresponds to the Hermite scheme developed by \citet{MakinoAarseth_1992_HermiteIntegratorAhmadCohen_PublicationsoftheAstronomicalSocietyofJapan}.
The first-time derivative of acceleration for the $i$-th particle within an NFW host potential is expressed as 
\begin{gather}
     \dot{\bm{a}}_i =-\sum_{j \neq i} G m_j\left[\frac{\bm{v}_{ij}}{r_{i j}^3}-\frac{3\left(\bm{r}_{ij} \cdot \bm{v}_{ij}\right) \bm{r}_{ij}}{r_{i j}^5}\right] - \nabla \dot{\Phi}_\mathrm{NFW,\,host}(r_i,v_i),
\end{gather}
where $\bm{v}_{ij}=\bm{v}_i - \bm{v}_j$, $v_{ij}=|\bm{v}_{ij}|$.
The time derivative of the force from the host halo is given by
\begin{gather}
    \nabla\dot{\Phi}_\mathrm{NFW,\,host}(r,v) = \frac{GM_\mathrm{200}g(c_\mathrm{host})}{g(x)}\left[\frac{\bm{v}}{r^3}-\frac{ (\bm{r}\cdot \bm{v})\bm{r}}{r^5}\left(3-\frac{x^2g(x)}{(1+x)^2}\right)\right],
\end{gather}
where $v=|\bm{v}|$ is the velocity of DMSHs.

In this paper, we adopt a shared time step for all DMSHs. The time step is determined by 
\begin{gather}
    \Delta t = \min_i\left(\Delta t_i\right),\quad \Delta t_i = \sqrt{\eta\frac{|\bm{a}_{1,i}||\bm{a}^{(2)}_{1,i}|+|\dot{\bm{a}}_{1,i}|^2}{|\dot{\bm{a}}_{1,i}||\bm{a}^{(3)}_{1,i}|+|\bm{a}^{(2)}_{1,i}|^2}}
\end{gather}
where $\bm{a}_{1,i}$ and $\dot{\bm{a}}_{1,i}$ are the acceleration and jerk calculated from predicted or corrected position and velocity of the $i$-th particle, respectively. The subscripts $0$ and $1$ indicate the value at $t$ and $t+\Delta t$, respectively.
$\bm{a}^{(2)}_{1,i}$ and $\bm{a}^{(3)}_{1,i}$ are given by
\begin{gather}
    \bm{a}^{(2)}_{1,i} = \bm{a}^{(2)}_{0,i} + \bm{a}^{(3)}_{0,i} \Delta t,\\
    \bm{a}^{(3)}_{1,i} = \bm{a}^{(3)}_{0,i},
\end{gather}
where 
\begin{gather}
    \bm{a}^{(2)}_{0,i}  =\frac{-6\left(\bm{a}_{0, i}-\bm{a}_{1, i}\right)-\Delta t_i\left(4 \dot{\bm{a}}_{0, i}+2 \dot{\bm{a}}_{1, i}\right)}{\Delta t_i^2},\\
    \bm{a}^{(3)}_{0,i} = \frac{12\left(\bm{a}_{0, i}-\bm{a}_{1, i}\right)+6 \Delta t_i\left(\dot{\bm{a}}_{0, i}+\dot{\bm{a}}_{1, i}\right)}{\Delta t_i^3}.
\end{gather}
Here, $\bm{a}_{0, i}$ and $\dot{\bm{a}}_{0, i}$ are calculated from the position and velocity of the $i$-th particle in the evaluation step.
The initial time step is calculated as
\begin{gather}
    \Delta t_\mathrm{init} = \min_i\left(\eta_\mathrm{s}\frac{|\bm{a}_i|}{|\dot{\bm{a}}_i|}\right).
\end{gather}
Two parameters of time step are set as $\eta = 0.005$ and $\eta_\mathrm{s} = 0.001$. 
We note that the gravitational softening parameter should be $0$ in this scheme, but to avoid the numerical divergence for the gravitational calculations, the small value $10^{-30}$ is added to the distance $r_{ij}$.

We run the orbital integration of DMSHs using the GPU-accelerated code implemented with OpenACC and Message Passing Interface \citep{OtakiEtAl_2024_AcceleratedHermiteIntegrator_}.
We analyze the relative distances and velocities of DMSH pairs using the output data in $0.1 \,\mathrm{Myr}$ intervals.
The collision events are defined as the pair of two DMSHs at the shortest relative distance and classified into three types depending on the relative distance of two DMSHs ($i$- or $j$-th DMSH) and their radius, such as
\begin{itemize}
    \item Violent encounter: 
    \begin{gather}
    r_{ij}\leq r_{\mathrm{s},\,i} + r_{\mathrm{s},\,j},
    \end{gather}
    \item Gentle encounter: 
    \begin{gather}
    r_{\mathrm{s},\,i} + r_{\mathrm{s},\,j} < r_{ij}\leq \min\left(r_{\mathrm{s},\,i}+R_{\mathrm{200},\,j},r_{\mathrm{s},\,j}+R_{\mathrm{200},\,i}\right),
    \end{gather}
    \item Grazing encounter: 
    \begin{gather}
    \min\left(r_{\mathrm{s},\,i}+R_{\mathrm{200},\,j},r_{\mathrm{s},\,j}+R_{\mathrm{200},\,i}\right) < r_{ij}\leq R_{\mathrm{200},\,i} + R_{\mathrm{200},\,j},
    \end{gather}
\end{itemize}
where $r_{ij}$ is the relative distance between two DMSHs and $r_\mathrm{s}$ and $R_\mathrm{200}$ are the scale radius and the virial radius of DMSHs.
To calculate the scale radius of DMSHs, we adopt the concentration derived from the maximum circular velocity for an NFW profile \citep{KlypinEtAl_2011_DARKMATTERHALOS_TheAstrophysicalJournal, PradaEtAl_2012_HaloConcentrationsStandard_MonthlyNoticesoftheRoyalAstronomicalSociety}.
The violent encounters of DMSHs are important events for the formation of stars and dwarf galaxies \citep{OtakiMori_2023_FrequencyDarkMatter_MonthlyNoticesoftheRoyalAstronomicalSociety}. 

\subsection{Results of simulations}
\begin{table}
\centering
\caption[Frequencies of encounters between dark matter subhalos in Milky Way-like host halos]{Frequencies of encounters between dark matter subhalos in Milky Way-like host halos: serial number of the host halo, frequencies of grazing encounters $f_\mathrm{grazing}$, gentle encounters $f_\mathrm{gentle}$ and violent encounters $f_\mathrm{violent}$, and fraction of the number of bound pairs to the total number of subhalos $f_\mathrm{bound}$.}
\label{tab: collision_count}
\tabcolsep=3pt
\begin{tabular}{ccccccc}
\hline
No. & $f_\mathrm{grazing}\,\mathrm{[Gyr^{-1}]}$ & $f_\mathrm{gentle}\,\mathrm{[Gyr^{-1}]}$ & $f_\mathrm{violent}\,\mathrm{[Gyr^{-1}]}$ & $f_\mathrm{bound}\,[\%]$ \\\hline
$1$ & $6.1 \times 10^4$ & $1.0 \times 10^4$ & $1.4 \times 10^2$  & $2.07$ \\
$2$ & $2.9 \times 10^4$ & $4.5 \times 10^3$ & $1.9 \times 10^2$  & $3.16$ \\
$3$ & $3.2 \times 10^4$ & $4.8 \times 10^3$ & $2.4 \times 10^2$  & $3.02$ \\
$4$ & $4.4 \times 10^4$ & $6.2 \times 10^3$ & $2.5 \times 10^2$  & $3.25$ \\
$5$ & $4.0 \times 10^4$ & $6.4 \times 10^3$ & $1.2 \times 10^2$  & $1.87$ \\
$6$ & $6.3 \times 10^4$ & $8.8 \times 10^3$ & $1.4 \times 10^2$  & $2.06$ \\
$7$ & $2.5 \times 10^4$ & $3.0 \times 10^3$ & $2.4 \times 10^2$  & $6.48$ \\
$8$ & $7.6 \times 10^4$ & $8.4 \times 10^3$ & $6.3 \times 10^2$  & $6.02$ \\
$9$ & $6.6 \times 10^4$ & $5.6 \times 10^3$ & $7.1 \times 10^2$  & $13.0$ \\
$10$ & $3.4 \times 10^4$ & $4.8 \times 10^3$ & $2.5 \times 10^2$  & $3.33$ \\
$11$ & $3.4 \times 10^4$ & $4.7 \times 10^3$ & $3.3 \times 10^2$  & $6.24$ \\
$12$ & $3.6 \times 10^4$ & $5.6 \times 10^3$ & $1.0 \times 10^2$  & $2.99$ \\
$13$ & $1.8 \times 10^4$ & $2.8 \times 10^3$ & $1.4 \times 10^2$  & $4.30$ \\
$14$ & $5.3 \times 10^4$ & $6.6 \times 10^3$ & $3.2 \times 10^2$  & $6.56$ \\
$15$ & $3.5 \times 10^4$ & $4.0 \times 10^3$ & $3.7 \times 10^2$  & $6.40$ \\
$16$ & $2.2 \times 10^4$ & $3.3 \times 10^3$ & $1.4 \times 10^2$  & $2.43$ \\
$17$ & $2.2 \times 10^4$ & $3.2 \times 10^3$ & $1.4 \times 10^2$  & $2.92$ \\
$18$ & $1.5 \times 10^4$ & $2.1 \times 10^3$ & $7.2 \times 10^1$  & $2.53$ \\
$19$ & $1.0 \times 10^4$ & $1.7 \times 10^3$ & $2.3 \times 10^1$  & $1.48$ \\
$20$ & $6.6 \times 10^4$ & $4.7 \times 10^3$ & $3.5 \times 10^2$  & $11.0$ \\
$21$ & $2.1 \times 10^4$ & $2.2 \times 10^3$ & $1.9 \times 10^2$  & $7.86$ \\
$22$ & $2.6 \times 10^4$ & $3.7 \times 10^3$ & $1.1 \times 10^2$  & $3.68$ \\
$23$ & $1.0 \times 10^4$ & $1.7 \times 10^3$ & $5.2 \times 10^1$  & $2.13$ \\
$24$ & $1.4 \times 10^4$ & $1.7 \times 10^3$ & $7.0 \times 10^1$  & $3.65$ \\
$25$ & $9.1 \times 10^3$ & $1.4 \times 10^3$ & $2.0 \times 10^2$  & $6.82$ \\
$26$ & $6.9 \times 10^3$ & $1.2 \times 10^3$ & $9.8 \times 10^1$  & $7.43$ \\
$27$ & $1.5 \times 10^4$ & $1.8 \times 10^3$ & $7.3 \times 10^1$  & $3.77$ \\

\hline
\end{tabular}
\end{table}

\begin{figure*}
    \centering
    \includegraphics[width=0.9\linewidth]{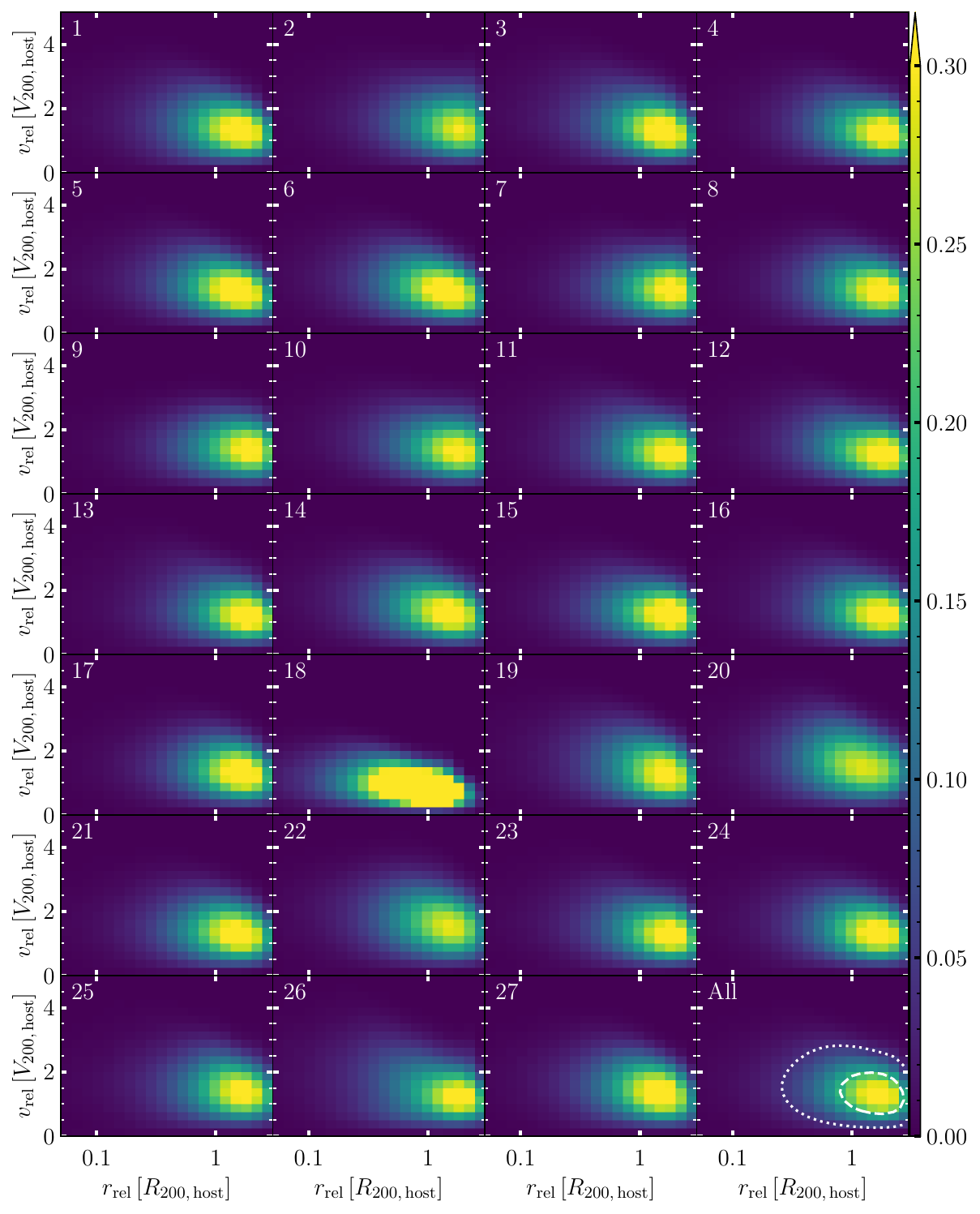}
    \caption{Density distribution of DMSHs encounters for relative distance $r_\mathrm{rel}$ and relative velocity $v_\mathrm{rel}$ between all DMSH pairs in each Milky Way-like host halo. The horizontal and vertical axes are normalised as the virial radius and the virial velocity of each host halo, respectively. Each panel is labelled with the number of the corresponding host halo. The bottom right panel shows the distribution stacked for all host haloes. The white dashed and dotted contours correspond to the regions that contain 39\% ($1\sigma$) and 86\% ($2\sigma$) of the total probability mass from the highest value probability density, respectively.}
    \label{fig: rrel_vrel_all}
\end{figure*}

\begin{figure*}
    \centering
    \includegraphics[width=0.9\linewidth]{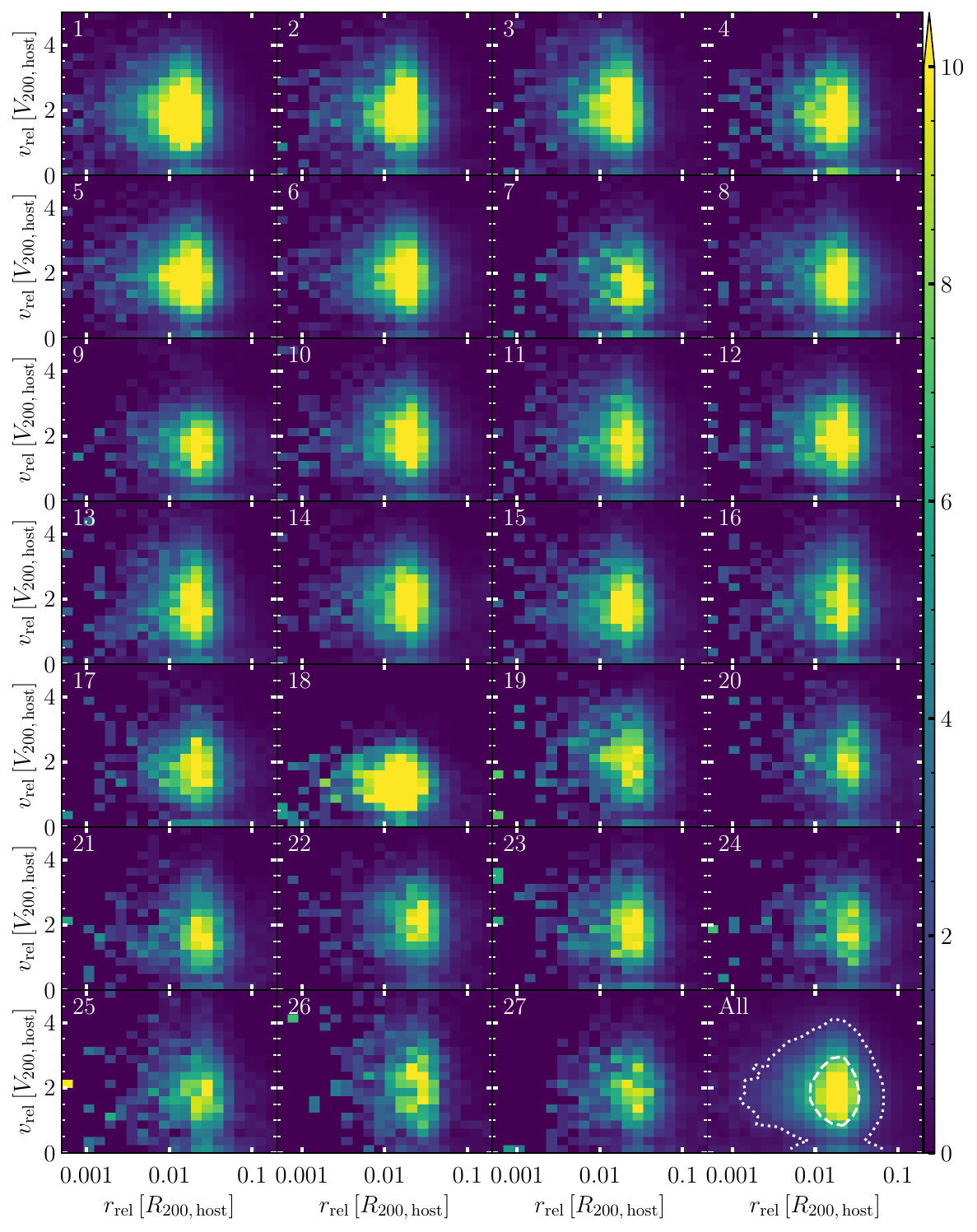}
    \caption{Same as Fig. \ref{fig: rrel_vrel_all}, but the density distribution for colliding DMSH pairs satisfied with the violent, gentle, and grazing encounters.}
    \label{fig: rrel_vrel}
\end{figure*}

\begin{figure*}
    \centering
    \includegraphics[width=0.9\linewidth]{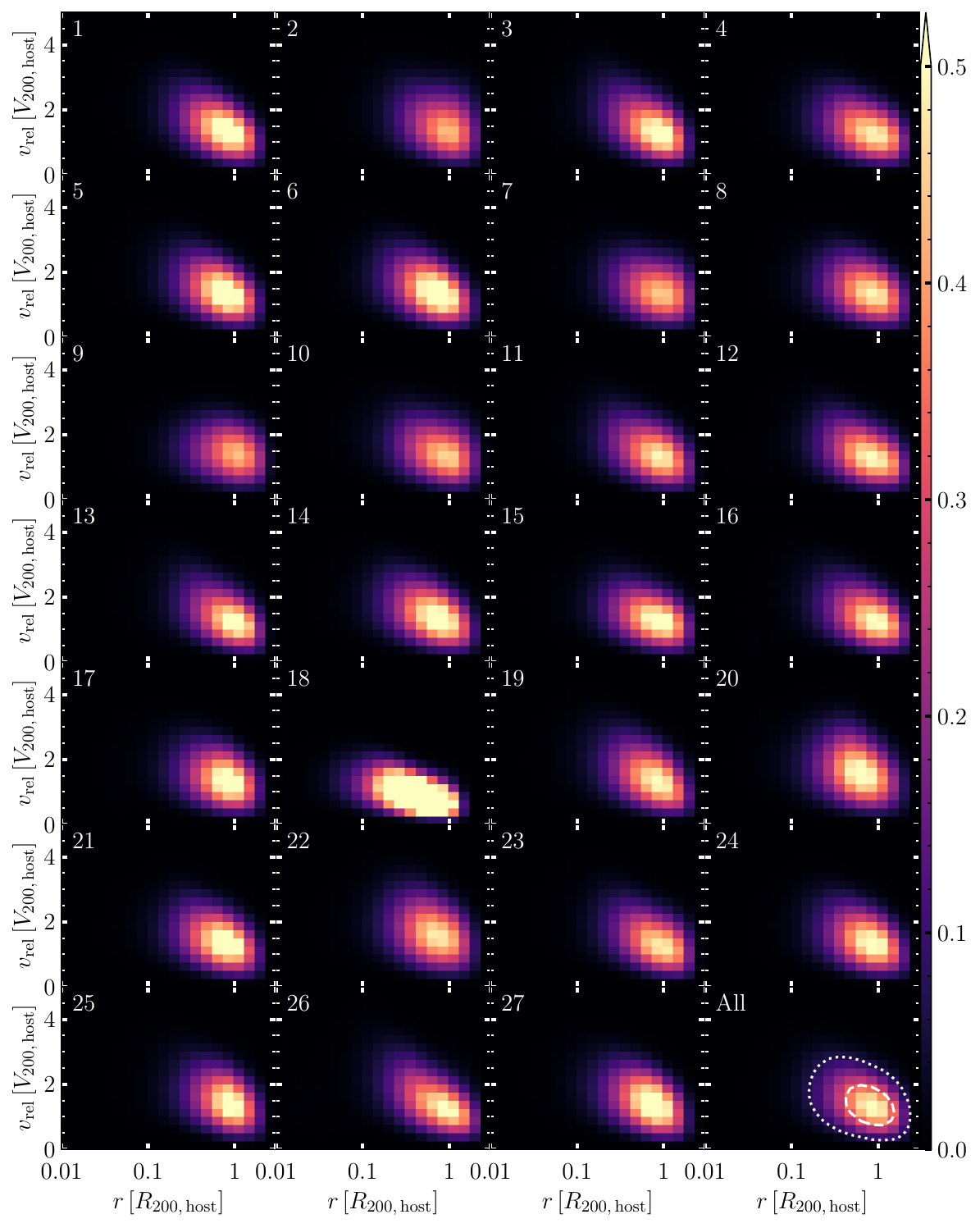}
    \caption{Density distribution of dark matter subhaloes encounters for the distance from the centre of each host halo $r$ and the relative velocity $v_\mathrm{rel}$ between colliding pairs of dark matter subhaloes in each Milky Way-like host halo. The horizontal and vertical axes are normalised as the virial radius and the virial velocity of each host halo, respectively. Each panel is labelled with the number of the corresponding host halo. The white dashed and dotted contours correspond to the regions that contain 39\% ($1\sigma$) and 86\% ($2\sigma$) of the total probability mass from the highest value probability density, respectively.}
    \label{fig: r_vrel_all}
\end{figure*}

\begin{figure*}
    \centering
    \includegraphics[width=0.9\linewidth]{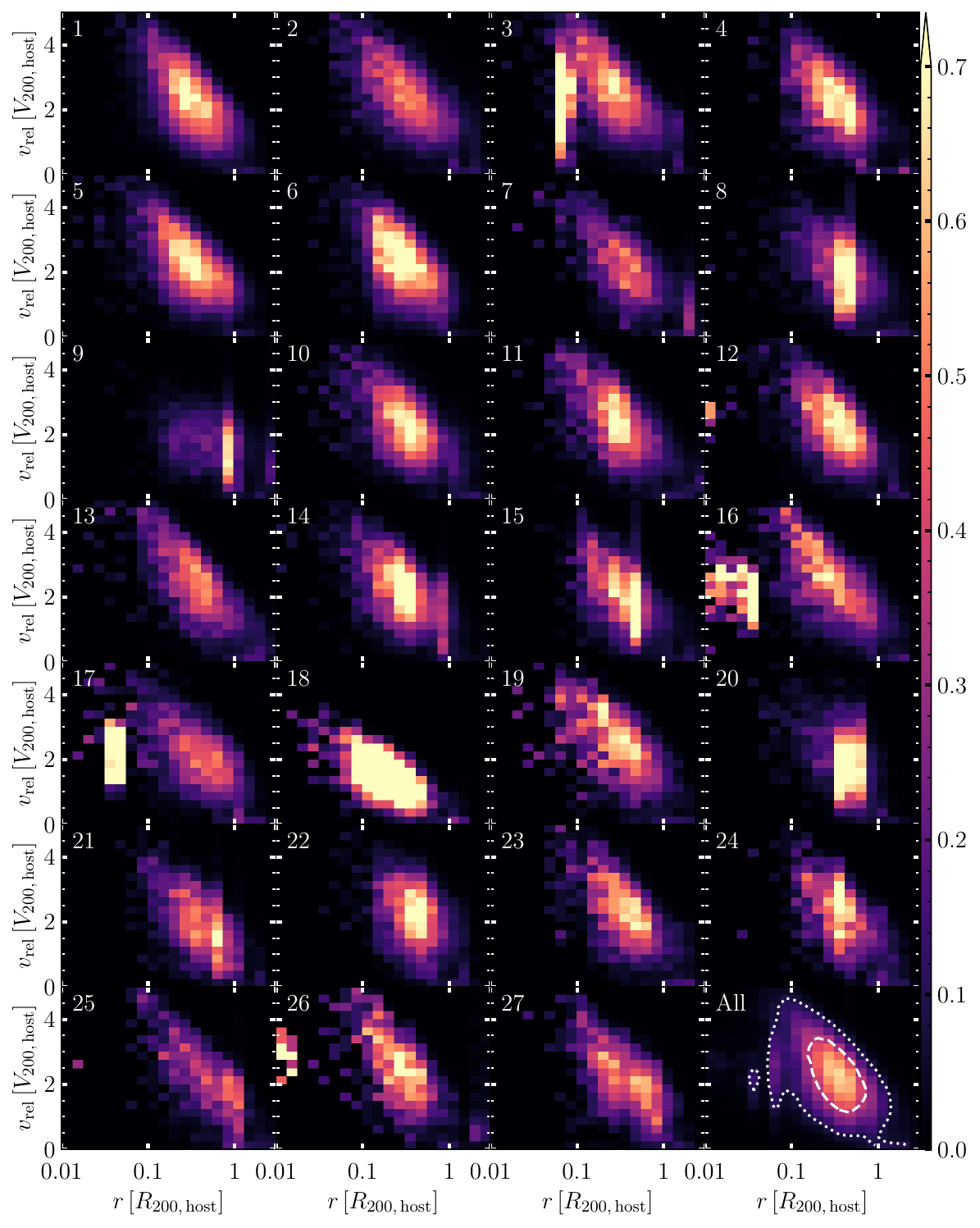}
    \caption{Same as Fig. \ref{fig: r_vrel_all}, but the density distribution for colliding DMSH pairs satisfied with the violent, gentle, and grazing encounters.}
    \label{fig: r_vrel}
\end{figure*}

We list the collision frequencies between DMSHs in Table \ref{tab: collision_count} for each MW-like host halo in the period from $\tau_\mathrm{sim}$ to $z = 0$.
Despite variations in the physical properties of each host halo, such as mass, concentration, and number of DMSHs, the classified collision frequencies are in almost the same order for all host haloes.
The average frequencies for all host haloes are $\overline{f}_\mathrm{grazing} = 3.3\times10^4\,\mathrm{Gyr^{-1}},\,\overline{f}_\mathrm{gentle} = 4.3\times10^3\,\mathrm{Gyr^{-1}},$ and $\overline{f}_\mathrm{violent} = 2.1\times10^2\,\mathrm{Gyr^{-1}}$.
In focusing on violent encounters with strong interaction, the time scale is calculated as $\tau_\mathrm{violent} = 4.7\,\mathrm{Myr}$, which is much shorter than the typical dynamical time of a host galaxy $R_{200,\,\mathrm{host}}/V_{200,\,\mathrm{host}}\sim1\,\mathrm{Gyr}$. 

Table \ref{tab: collision_count} mentions the numbers and fractions of gravitationally bound pairs satisfied with the negative energy such as $E_{ij}=v_{ij}^2/2 - G(m_i + m_j)/r_{ij} < 0$, where $v_{ij} = |\bm{v}_i - \bm{v}_j|$, and $r_{ij} = |\bm{r}_i - \bm{r}_j|$ for the pairs between $i$- and $j$-th DMSHs.
The fraction of bound pairs is at least several percent, which indicates that there could be gravitationally strongly interacting binary DMSHs in galaxies.
These pairs could merge and form a single DMSH. In the case of a high baryonic fraction within DMSHs, a dwarf galaxy will be formed after a merger \citep{OtakiMori_2023_FrequencyDarkMatter_MonthlyNoticesoftheRoyalAstronomicalSociety}.

Fig. \ref{fig: rrel_vrel_all} shows the density distributions of all DMSH pairs in each MW-like host halo.
The vertical and horizontal axes are the relative distances and velocities normalised by the virial radius and the circular velocity at the virial radius for each host halo, respectively, when the pair approaches closest to each other within the orbital integration time.
The colour represents the number density of DMSH pairs per bin area derived as $N_\mathrm{pair}(r_\mathrm{rel},\,v_\mathrm{rel})/(N_\mathrm{pair,\, tot}\,\Delta r_\mathrm{rel}\,\Delta v_\mathrm{rel})$.
The lower right panel in Fig. \ref{fig: rrel_vrel_all} indicates the stacked distribution of numbers of DMSH pairs for all host haloes.
The stacked distribution reveals that the peak positions of the relative distance and relative velocity are $1.6 \,R_\mathrm{200,\, host}$ and $1.3 \,V_\mathrm{200,\, host}$, respectively. 
Numerical results suggest that the distributions are similar to the analytical model (Fig. \ref{fig: Prel}) in tendency whereas the peak positions of the relative distance and relative velocity are smaller than those of the analytical model. 
We note that the analytical model ignores the interaction between DMSHs and the mass function of the DMSH.
While the peak position of the relative distance, the $r_\mathrm{rel}$-$v_\mathrm{rel}$ distribution of DMSH could be represented by the phase distribution function of the host halo.

Instead of Fig. \ref{fig: rrel_vrel_all}, the selected density distributions by colliding DMSH pairs classified as the violent, gentle, and grazing encounters are shown in Fig. \ref{fig: rrel_vrel}. These pairs are satisfied with the condition of $r_{ij}\leq R_{200,\,i}+R_{200,\,j}$.
The colour denotes the normalised density distribution for all DMSH collisions, expressed as $N_\mathrm{col}(r_\mathrm{rel},\,v_\mathrm{rel})/(N_\mathrm{col,\,tot}\,\Delta  r_\mathrm{rel}\,\Delta v_\mathrm{rel})$, where $N_\mathrm{col}(r_\mathrm{rel},\,v_\mathrm{rel})$ is the number of collisions between DMSHs with $r_\mathrm{rel}$ and $v_\mathrm{rel}$ including grazing, gentle, and violent encounters, and $N_\mathrm{col,\, tot}$ is the total number of collisions in the host halo.
Although the density distributions for each host halo are somewhat noisy due to the small sample sizes of the collision pairs, they are consistent across host haloes regarding the peak positions of the relative distances and velocities, as well as in their overall trends.
The stacked density distribution for all haloes is approximately symmetrical around the peak positions of the relative distance and velocity, at $0.019 \,R_\mathrm{200,\, host}$ and $1.8 \,V_\mathrm{200,\, host}$ respectively. Notably, there are minor distributions within $0.01 \,R_{200,\,\mathrm{host}}\lesssim r \lesssim 0.1 \, R_{200,,\mathrm{host}}$, indicative of low-velocity collisions.
It also shows that gravitationally bound pairs with short relative distances occur in the host halo.

Figs. \ref{fig: r_vrel_all} and \ref{fig: r_vrel} represent the density distribution of all pairs and colliding pairs of DMSHs for the distance from the centre of the host halo and the relative velocity in the period from $\tau_\mathrm{sim}$ to $z = 0$, such as 
$N_\mathrm{pair}(r,\,v_\mathrm{rel})/(N_\mathrm{pair,\,tot}\,\Delta  r\,\Delta v_\mathrm{rel})$ and $N_\mathrm{col}(r,\,v_\mathrm{rel})/(N_\mathrm{col,\,tot}\,\Delta  r\,\Delta v_\mathrm{rel})$, respectively.
The peak positions of the relative distance in the stacked distribution are $0.79 \,R_\mathrm{200,\, host}$ and $0.33 \,R_\mathrm{200,\, host}$ for all and colliding pairs, respectively.
The DMSH collision events occur in the inner region of the host halo. 
Although Fig. \ref{fig: r_vrel} suggests that the distributions for colliding DMSHs vary locally based on the properties of the host halo, globally, these distributions show a trend where the fraction of collisions with higher relative velocities increases closer to the center of the host haloes.
Conversely, the relative velocities of pairs colliding near the virial radius diminish as the velocities of the DMSHs decrease. This leads to the distributions of low-velocity collisions, approximately $v_\mathrm{rel}\sim0$, observed in Fig. \ref{fig: rrel_vrel}, which are attributed to collisions at around the virial radius, $r\sim R_{200,\,\mathrm{host}}$.

Compared to Fig.~\ref{fig: col_freq}, which is derived from the analytical model for DMSH collision frequency, the numerical results exhibit similar distributions and peak locations for position and relative velocity. Specifically, the peak values in the analytical model are $r_\mathrm{peak}=0.310\,R_{200,\,\mathrm{host}}$ and $v_\mathrm{rel,\,peak}=1.43\,V_{200,\,\mathrm{host}}$, while those in the numerical results are $0.33\,R_{200,\,\mathrm{host}}$ and $1.8\,V_{200,\,\mathrm{host}}$, respectively. However, the slightly higher probability of relative velocities at the higher end in Fig.~\ref{fig: r_vrel} can be attributed to the effects of self-gravity between DMSHs, which are not incorporated in the analytical model. Furthermore, tidal interactions induced by the gravitational potential of the host system play a significant role in disrupting DMSHs, leading to a reduction in their number density within the inner regions of the halo.

\section{Summary and Discussion} \label{sec: sumdis}
We have derived the collision frequency distribution of DMSHs in MW-like host haloes using analytical and numerical studies.
Our two analytical models derived from the distribution function of the host halo show the $r_\mathrm{rel}$-$v_\mathrm{rel}$ distribution for all DMSH pairs (Fig. \ref{fig: Prel}) and the $r$-$v_\mathrm{rel}$ distribution for colliding DMSH pairs (Fig. \ref{fig: col_freq}).
The $r_\mathrm{rel}$-$v_\mathrm{rel}$ distribution indicates a relatively low probability for DMSH pairs with relative distances of $\lesssim 0.1\, R_{200,\,\mathrm{host}}$ or relative velocities of $\gtrsim 3 V_{200,\,\mathrm{host}}$.
Compared to \citet{OtakiMori_2023_FrequencyDarkMatter_MonthlyNoticesoftheRoyalAstronomicalSociety}, the $r$-$v_\mathrm{rel}$ distribution, which includes the number density of DMSHs and accounts for tidal disruption, suggests that DMSH collision frequencies are lower in the inner region ($r\lesssim 0.1 R_\mathrm{200,\,\mathrm{host}}$).

Then, we simulate the orbital evolution of the DMSHs in the Milky Way-like host halo and analyse the collision frequencies between DMSHs classified by grazing, gentle, and violent encounters for the virial and scale radii of DMSHs.
We derive that the average frequencies are $\overline{f}_\mathrm{grazing} = 3.3\times10^4\,\mathrm{Gyr^{-1}},\,\overline{f}_\mathrm{gentle} = 4.3\times10^3\,\mathrm{Gyr^{-1}},$ and $\overline{f}_\mathrm{violent} = 2.1\times10^2\,\mathrm{Gyr^{-1}}$, and the collision time scale of violent encounters is estimated as $\tau_\mathrm{violent} = 4.7\,\mathrm{Myr}$. 
The violent encounters could induce bursts of star formation and lead to the formation of dwarf galaxies as this time scale is shorter than the typical dynamical time of a host galaxy $R_{200,\,\mathrm{host}}/V_{200,\,\mathrm{host}}\sim1\,\mathrm{Gyr}$.
We found that the numerical results were similar to the analytical models for DMSH pair density distributions, despite the analytical model neglecting interactions between DMSHs.
It suggests the properties of DMSH collision are approximately characterised by the host halo.
However, the gravitational interactions between DMSHs produce DMSH pairs with a high relative velocity within the inner region of the host halo $r\lesssim 0.1\, R_{200,\,\mathrm{host}}$ compared to the analytical model.

Since we selected DMSHs identified at redshift $z=0$ from the result of the cosmological simulation Phi-4096 \citep{IshiyamaEtAl_2021_UchuuSimulationsData_MonthlyNoticesoftheRoyalAstronomicalSociety}, the frequency of violent encounters may be slightly underestimated due to the exclusion of 
pre-merged DMSHs at $z = 0$. Collision simulations based on merger histories can include the evolution of the collisions between DMSHs before merging at $z = 0$. Then, the number of DMSH collisions may effectively increase more than in this study, leading to a shorter collision timescale. In addition,
the orbital integration time is set to several hundred Myr to minimise the impact of the changes in the mass evolution of host haloes. 
Therefore, a discussion about the evolution of collision frequencies including merged DMSHs remains a subject for a future study.
In the following, we compare our results with observations. Then we discuss the fraction of dark satellite interactions from our results.

\begin{figure*}
    \centering
    \includegraphics[width=\linewidth]{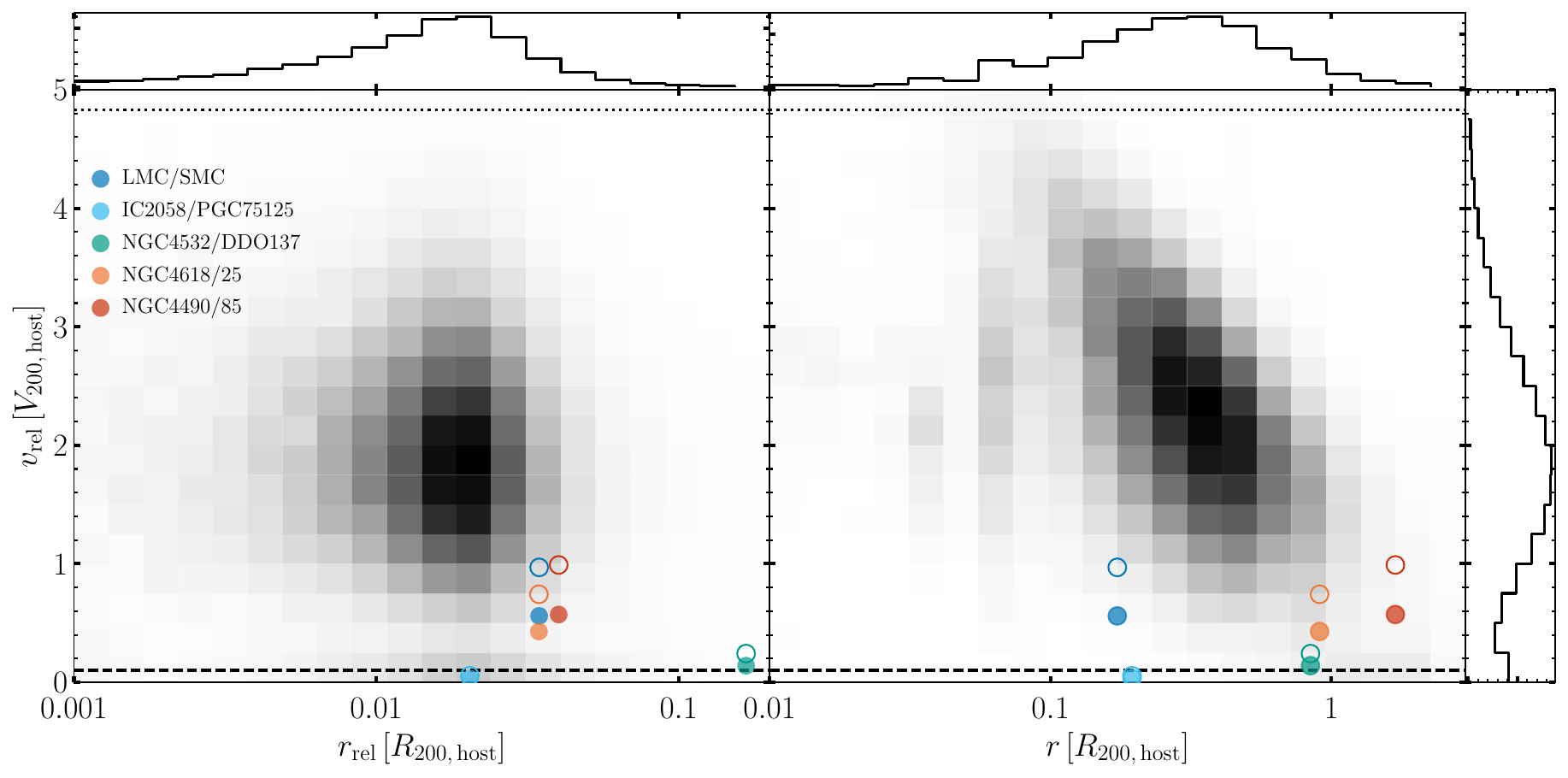}
    \caption{Comparison between observations and results of collision simulations. The background of the left and right panels shows the density distribution of stacked dark matter subhalo collisions from the simulations in the $r_\mathrm{rel}$-$v_\mathrm{rel}$ plane and $r$-$v_\mathrm{rel}$-plane, respectively. The filled circles represent the properties of the observed satellite galaxy pairs \citep{PearsonEtAl_2016_LocalVolumeTiNy_MonthlyNoticesoftheRoyalAstronomicalSociety}. The open circles show relative velocities corresponding to the line-of-sight velocities multiplied by $\sqrt{3}$.
    The dashed and dotted lines indicate the critical relative velocities for the merging and the occurring shock-breakout via a head-on collision of two dark matter subhaloes with $10^9\,\mathrm{M_\odot}$, respectively \citep{OtakiMori_2023_FrequencyDarkMatter_MonthlyNoticesoftheRoyalAstronomicalSociety}.
    }
    \label{fig: obs}
\end{figure*}

\subsection{Comparison to observed dwarf-dwarf interactions}
We compare our results with observations and discuss the cold dark matter paradigm in the aspect of the DMSH collisions.
\citet{PearsonEtAl_2016_LocalVolumeTiNy_MonthlyNoticesoftheRoyalAstronomicalSociety} listed the dynamical properties of observed dwarf-dwarf interactions in the Local Universe. Their sample includes six interacting pairs associated with each host halo, LMC/SMC, IC2058/PGC75125, NGC4532/DDO137, NGC4618/25, NGC4490/85, and ESO435-IG16/IG20.
We note that the pair of ESO435-IG16/IG20 listed as satellites in \citet{PearsonEtAl_2016_LocalVolumeTiNy_MonthlyNoticesoftheRoyalAstronomicalSociety} is removed in this discussion because it is located more distant than twice the virial radius of its host galaxy.
The host halo masses with observed pairs are in the range from $\sim7\times10^{11}\,\mathrm{M_\odot}$ to $\sim 1\times10^{13}\,\mathrm{M_\odot}$.

Fig. \ref{fig: obs} shows the comparison between the observational data and our simulation results. 
The left panel shows the $r_\mathrm{rel}$-$v_\mathrm{rel}$ diagram for the dwarf galaxy pair, and the right panel shows the distance from the centre of the host galaxy and relative velocity. The background distributions show the stacked results of our simulation for all MW-like haloes.
Since the observed galaxy pairs represent line-of-sight velocities, the data points multiplied by $\sqrt{3}$ are also included in Fig. \ref{fig: obs} as open circles.
The left and right upper sub-panels correspond to the distributions of DMSH pairs of the relative distance and the distance from the centre of the host halo, respectively. The right sub-panel indicates the relative velocity distribution of DMSH pairs.
Although the small number of observed samples limits the statistical arguments, the numerical results indicate that DMSH pairs have relatively high-velocity interactions compared to observed galaxy pairs.

In Fig. \ref{fig: obs}, we represent dashed and dotted lines corresponding to the critical relative velocities of merger and shock-breakout conditions by gas-rich DMSH collisions modelled by \citet{OtakiMori_2023_FrequencyDarkMatter_MonthlyNoticesoftheRoyalAstronomicalSociety}, respectively.
In the case of a low-speed collision in which the total energy is negative, two colliding DMSHs merge into a single gravitationally bound system.
The critical relative velocity satisfied with the merger condition is $v_\mathrm{merger}=14.5\,\mathrm{km\,s^{-1}}$ for the DMSH mass of $10^9\,\mathrm{M_\odot}$.
On the other hand, a high-speed collision induces the shock-breakout due to the shock wave generated at the collision surface. Then most of the gas in DMSHs is ejected and no galaxy forms. \citet{OtakiMori_2023_FrequencyDarkMatter_MonthlyNoticesoftheRoyalAstronomicalSociety} modelled the one-dimensional gas cloud collision to derive the shock-crossing time $t_\mathrm{cross}$ and compare it with the gas cooling time $t_\mathrm{cool}$. For $t_\mathrm{cool}<t_\mathrm{cross}$, the effective radiative cooling enhances the star formation. In the case of $t_\mathrm{cross} < t_\mathrm{cool}$, the shock-breakout occurs and suppresses the star formation. 
The critical relative velocity of $t_\mathrm{cross} = t_\mathrm{cool}$ is calculated as $v_\mathrm{crit}=691\,\mathrm{km\,s^{-1}}$ for the DMSH mass of $10^9\,\mathrm{M_\odot}$. 
The DMSH collision satisfying the relative velocity condition $v_\mathrm{merger}<v_\mathrm{rel}<v_\mathrm{cross}$ results in the formation of a dark-matter-deficient galaxy \citep[see][for more details]{OtakiMori_2023_FrequencyDarkMatter_MonthlyNoticesoftheRoyalAstronomicalSociety}.

The numerical results suggest a higher probability of observing galaxy pairs with high-velocity interactions. Such pairs are unlikely to be observed as dwarf mergers, as they may instead form dark-matter-deficient galaxies or fail to form galaxies altogether due to shock-breakout.
Additionally, these collision events are difficult to observe because their interaction timescale is comparable to the dynamical time.
In recent theoretical studies, the collision process between dwarf galaxies and DMSHs induces the formation of dark-matter-deficient galaxies \citep{Silk_2019_UltradiffuseGalaxiesDark_MonthlyNoticesoftheRoyalAstronomicalSociety:Letters, LeeEtAl_2021_DarkMatterDeficient_TheAstrophysicalJournalLetters, OtakiMori_2023_FrequencyDarkMatter_MonthlyNoticesoftheRoyalAstronomicalSociety, LeeEtAl_2024_MultipleBeadsString_TheAstrophysicalJournal}.
Furthermore, advances in observational technology have led to the discovery of numerous faint galaxies in the Milky Way \citep[e.g.,][]{HommaEtAl_2024_FinalResultsSearch_PublicationsoftheAstronomicalSocietyofJapan}.
As part of the Missing Satellite problem, additional faint dark galaxies may be detected in future observations.

\subsection{Fraction of dark satellites in Milky Way-like haloes}

We provide an estimate of the fraction of observed bright satellites relative to the number of DMSHs in a host halo, based on observations of dwarf-dwarf interactions.
In the following, both bright and dark dwarf galaxies are collectively referred to as DMSHs.

The fraction of bright satellites is given as $f_\mathrm{bright} = N_\mathrm{bright}/N_\mathrm{sub}$, where $N_\mathrm{bright}$ and $N_\mathrm{sub}$ are the number of bright satellites and total DMSHs in the host halo, respectively. 
We assume that the number of collisions per each DMSH can also be written as $f_\mathrm{bright}=M_\mathrm{bright}/M_\mathrm{sub}$, where $M_\mathrm{bright}$ and $M_\mathrm{sub}$ are the collision counts per DMSH for other bright satellites and DMSHs, respectively.
Using the collision frequency $f_\mathrm{col}$ derived from our numerical simulations, 
the total number of collisions between the DMSHs within a host halo is expressed as $f_\mathrm{col}\Delta t =  N_\mathrm{sub}M_\mathrm{sub} / 2=N_\mathrm{bright}M_\mathrm{bright} / (2f_\mathrm{bright}^2)$. Here $\Delta t$ represents the timescale on which the collision signs remain without dynamical relaxation.
In the case of the Milky Way, the number of bright satellites is $N_\mathrm{bright}\simeq 50$ and the number of collisions between satellites is $N_\mathrm{bright}M_\mathrm{bright}/2=1$, since there is only LMC-SMC pair.
Therefore, the fraction of bright satellites is given by
\begin{gather}
    f_\mathrm{bright} \simeq 0.2\left(\frac{f_\mathrm{col}}{200,\,\mathrm{Gyr^{-1}}}\right)^{-1/2}\left(\frac{\Delta t}{100\,\mathrm{Myr}}\right)^{-1/2}\left(\frac{N_\mathrm{bright}}{50}\right)^{1/2}\left(\frac{M_\mathrm{bright}}{0.04}\right)^{1/2},
\end{gather}
where we assume a typical dynamical timescale of dwarf galaxies $100\,\mathrm{Myr}$ as $\Delta t$.
In other words, the fraction of bright-bright, bright-dark, and dark-dark collisions are given by
\begin{gather}
    f_\mathrm{bright\text{-}bright} = f_\mathrm{bright}f_\mathrm{bright} \simeq 0.05,\\
    f_\mathrm{bright\text{-}dark} = 2f_\mathrm{bright}f_\mathrm{dark} \simeq 0.35,\\
    f_\mathrm{dark\text{-}dark} = f_\mathrm{dark}f_\mathrm{dark} \simeq 0.60,
\end{gather}
respectively. Here, the fraction of dark satellites is represented as $f_\mathrm{dark} = 1 - f_\mathrm{bright}$.
Thus, the ratio between bright-dark collisions and bright-bright collisions is expressed by
\begin{gather}
    \frac{f_\mathrm{bright\text{-}dark}}{f_\mathrm{bright\text{-}bright}} = \frac{2f_\mathrm{bright}f_\mathrm{dark}}{f_\mathrm{bright}^2} \simeq 7.
\end{gather}
Collisions between bright and dark satellites, as well as between dark satellites, are significantly more frequent than observed collisions between bright galaxies.
Future observations of fainter structures as signatures of DMSH collisions could enable us to estimate their number and frequency.
We expect that signatures of dwarf galaxy interactions could provide evidence for the existence of dark galaxies that have yet to be discovered within the CDM paradigm.
For alternative dark matter models, investigating the frequency of DMSH collisions and their associated processes is crucial for constraining these models.

Finally, we consider the spherical potential of the MW-like host halo, but the triaxiality of the halo is a factor in the spatial distribution of the DMSHs \citep[e.g.,][]{Hoffmann_2014}. Our results provide a good approximation for collisions near the centre of the halo, while the outer regions may be more sensitive to the triaxial structure and require further investigation. Therefore, accounting for the impact of triaxial anisotropy of the host halo on the collision frequency distribution of DMSHs could provide important insights into the observed fraction of colliding dwarf galaxies. 

\section*{Acknowledgements}
We are very grateful to Tomoaki Ishiyama for providing the results of the cosmological $N$-body simulation Phi-4096. {We thank Yohei Miki and Takanobu Kirihara for fruitful discussions. We also thank the anonymous referee for helpful suggestions.}
This work was supported by JSPS KAKENHI Grant Numbers P20K04022, JP22KJ0370, JP24K07085 and JP24K00669.
Numerical computations were performed with computational resources, Cygnus and Pegasus provided by the Multidisciplinary Cooperative Research Program in the Center for Computational Sciences, the University of Tsukuba and the GPU system at the Center for Computational Astrophysics, National Astronomical Observatory of Japan.
KO acknowledges support from the PRIN 2022 MUR project 2022CB3PJ3—First Light And Galaxy aSsembly (FLAGS) funded by the European Union—Next Generation EU, and from EU-Recovery Fund PNRR - National Centre for HPC, Big Data and Quantum Computing.
\appendix

\section{Formulation of concentration-mass relation}\label{app:c-M}

We adopt the $c\text{-}M$ relation proposed by \citet{DiemerJoyce_2019_AccuratePhysicalModel_TheAstrophysicalJournal},
\begin{gather}
    c=C\left(\alpha_{\mathrm{eff}}\right) \times \tilde{G}\left(\frac{A\left(n_{\mathrm{eff}}\right)}{\nu}\left[1+\frac{\nu^2}{B\left(n_{\mathrm{eff}}\right)}\right]\right),
\end{gather}
where $\alpha_\mathrm{eff}$ and $n_\mathrm{eff}$ are the effective slope of the linear growth rate of fluctuations and root mean square density fluctuation $\sigma(M)$ expressed by 
\begin{gather}
    n_{\mathrm{eff}}(M)=-\left.2 \frac{\mathrm{d} \ln \sigma(R)}{\mathrm{d} \ln R}\right|_{R=\kappa R_{\mathrm{L}}}-3,
\end{gather}
and 
\begin{gather}
    \alpha_{\mathrm{eff}}(z)=-\frac{\mathrm{d} \ln D(z)}{\mathrm{d} \ln (1+z)},
\end{gather}
respectively.
$\sigma(R)$ is the rms density fluctuation in spheres of Lagrangian radius $R_\mathrm{L}$ and $D(z)$ is the linear growth factor.
The Lagrangian radius is defined as
\begin{gather}
    M = \frac{4\pi}{3}\rho_\mathrm{m}R_\mathrm{L}^3,
\end{gather}
where $M$ is the halo mass and $\rho_\mathrm{m}$ is the mean density of the universe at redshift $z=0$.
$\nu = \delta_\mathrm{c}/\sigma(M)$ is the peak height of density and $\delta_\mathrm{c}=1.686$ is the critical overdensity for the spherical collapse.
$\tilde{G}(x)$ is the inverse function of 
\begin{gather}
    G(x)=\frac{x}{[f(x)]^{\left(5+n_{\mathrm{eff}}\right) / 6}},
\end{gather}
The functions of $A(n_\mathrm{eff}), \, B(n_\mathrm{eff}),$ and $C(\alpha_\mathrm{eff})$ are given by
\begin{align}
A\left(n_{\mathrm{eff}}\right) & =a_0\left(1+a_1\left(n_{\mathrm{eff}}+3\right)\right), \\
B\left(n_{\text {eff }}\right) & =b_0\left(1+b_1\left(n_{\mathrm{eff}}+3\right)\right), \\
C\left(\alpha_{\mathrm{eff}}\right) & =1-c_\alpha\left(1-\alpha_{\mathrm{eff}}\right).
\end{align}
This formula of the $c$-$M$ relation is parameterised by six free parameters, $\kappa,\,a_0,\,a_1,\,b_0,\,b_1$ and $c_\alpha$. 
In this paper, we adopt $\kappa=1.10,\,a_0=2.30,\,a_1=1.64,\,b_0=1.72,\,b_1=3.60,$ and $c_\alpha=0.32$ provided by \citet{IshiyamaEtAl_2021_UchuuSimulationsData_MonthlyNoticesoftheRoyalAstronomicalSociety}.

\section{Parameter dependence in the analytical model} \label{app:anal}
\begin{figure}
    \centering
    \includegraphics[width=\columnwidth]{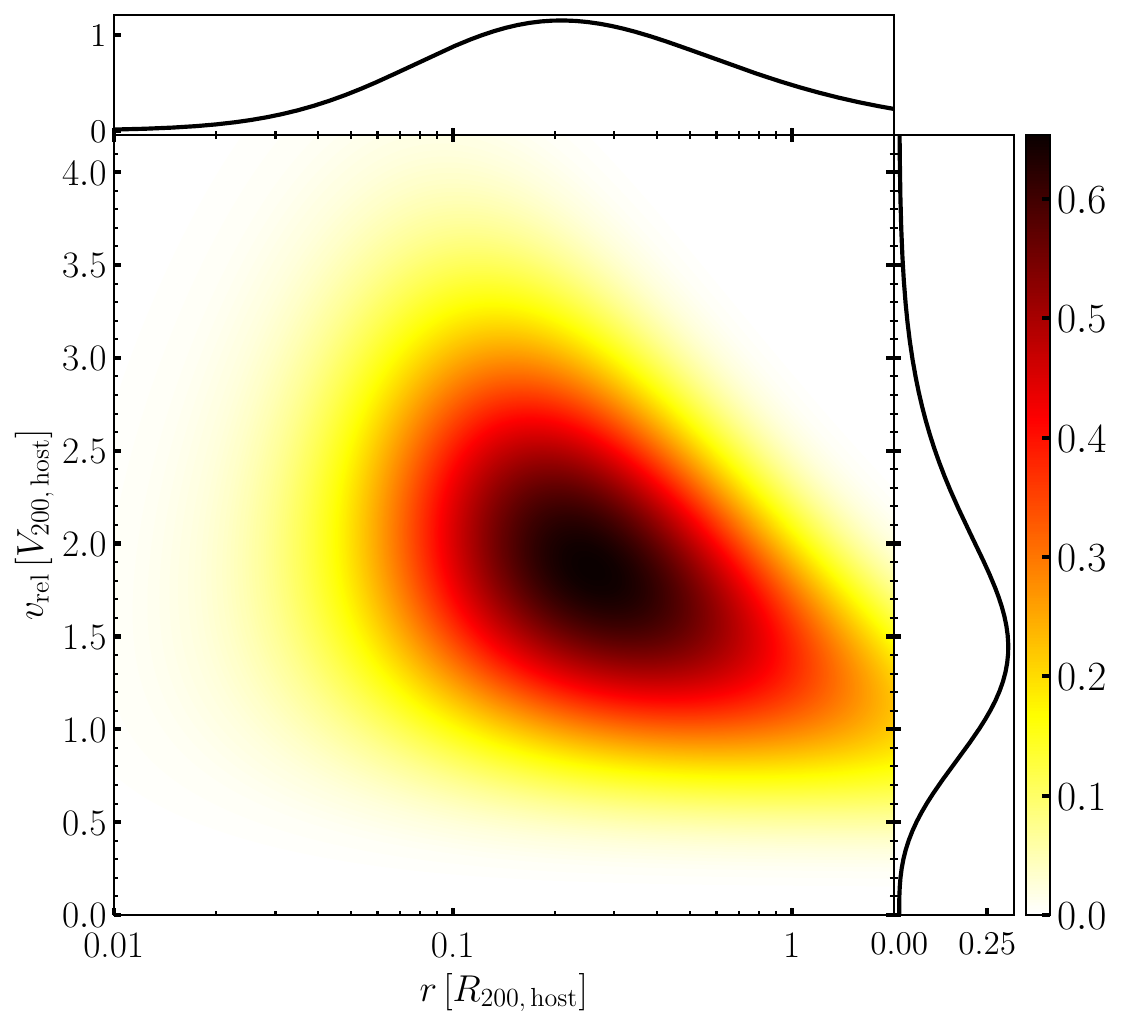}
    \caption{Same as Fig. \ref{fig: Prel}, but for the concentration of the host halo with $c_\mathrm{host}=12.0$.}
    \label{fig: Prel_c12}
\end{figure}

In our analytical model, we adopt the concentration parameter of a host halo, $c_\mathrm{host}=8.14$, and the slope of the spatial distribution of DMSHs, $\gamma=1.5$ as the fiducial parameters. The parameters reflect the properties of individual host halos and DMSH populations. Here, we demonstrate the two results of the analytical model with different parameters.
Fig. \ref{fig: Prel_c12} shows the distribution of collision frequency ${c_\mathrm{sub}^2\mathrm{d}k_\gamma}/({N_\mathrm{sub}^2\eta^2\mathrm{d} r\,\mathrm{d}v_\mathrm{rel}\,\mathrm{d}t})$ for $\gamma = 1.5$ and $c_\mathrm{host}=12.0$.
Due to the higher concentration of the host halo than the fiducial model shown in Fig. \ref{fig: Prel}, the overall distribution shifts slightly towards the centre of the halo, but the distribution of the relative velocity is not changed by the value of the concentration. The collision events occur most frequently at the position of $r_\mathrm{peak}=0.208\,R_{200,\,\mathrm{host}}$ with the relative velocity $v_\mathrm{rel,\,peak}=1.43\,V_{200,\,\mathrm{host}}$. 

\begin{figure}
    \centering
    \includegraphics[width=\columnwidth]{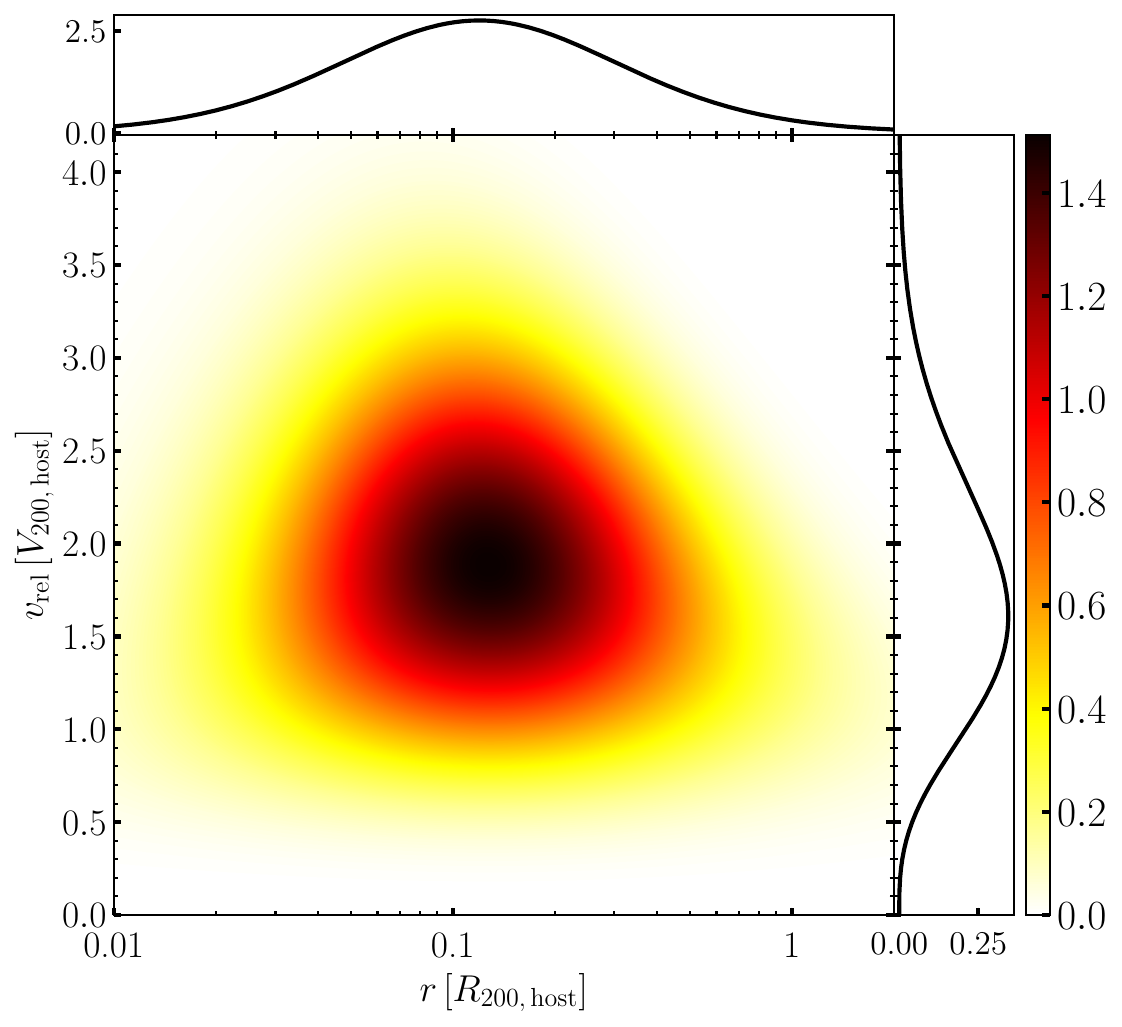}
    \caption{Same as Fig. \ref{fig: Prel}, but for the power exponent of the spatial distribution of DMSHs with $\gamma=1$.}
    \label{fig: Prel_g1}
\end{figure}

In this study, we derived the average value $\gamma=1.5$ based on the spatial distribution of DMSHs with $M_\mathrm{vir}>10^6\,\mathrm{M_\odot}$. However, when we select the DMSHs with $M_\mathrm{vir}>10^7\,\mathrm{M_\odot}$, the parameter becomes $\gamma=1.1$ on average due to the low number density in the central regions of host haloes compared to the fiducial selection of DMSHs.
Here, we show a distribution of collision frequency for the model with $\gamma = 1$ and $c_\mathrm{host}=8.14$ in Fig. \ref{fig: Prel_g1} to compare with Fig. \ref{fig: Prel}. For smaller $\gamma$, the spatial distribution of DMSHs becomes similar to the NFW profile. In this case, the number density profile in the central and outer regions is $\propto r^0$ and $\propto r^{-2}$, respectively. Compared to $\gamma=1.5$, the slope of the number density in the outer regions is steeper, resulting in collisions occurring more frequently in the central region of the host halo. Consequently, the shape of the distribution of the collision frequency depends on the spatial distribution of DMSHs.
The peak position and relative velocity of the most frequent collisions are $r_\mathrm{peak}=0.119\,R_{200,\,\mathrm{host}}$ and $v_\mathrm{rel,\,peak}=1.61\,V_{200,\,\mathrm{host}}$, respectively.
If positions where collision events frequently occur could be determined through observations, it may be possible to infer the spatial distribution of DMSH based on our analytical model.

\section{Effects of dark matter subhalo properties on collision frequencies}\label{app:col}
The collision frequency in a host halo depends on the DMSH properties, such as mass and infall time. 
Here, we investigate the effects of the most massive DMSHs and the first infalling and multiple orbiting DMSHs.

The Milky Way-like host halo (No. 9) has the most massive DMSH with the highest mass fraction for the host halo compared to the other models. The mass of the most massive DMSH is $2.2\times10^{11}\,\mathrm{M_\odot}$, corresponding to a mass fraction of 27.3\%.
The contributions of this DMSH to the violent, gentle, and grazing encounters relative to the total encounters are 29\%, 0.7\%, and 12\%, respectively.
Since the massive DMSHs have large radii and deep gravitational potentials, close encounters with the massive DMSHs could occur more frequently compared to the less massive DMSHs. Due to the massive DMSHs affecting the dynamics within a host halo, dependent analysis of the collision frequencies on DMSH masses could provide insights into collision events in observations to inspect the $\Lambda$CDM model.
Since massive DMSHs affect the dynamics within host haloes, analysing the collision frequency dependence on DMSH mass could provide the effect on collision dynamics between DMSHs.

We define the ``first-infall'' DMSHs as those that were accreted approximately one dynamical time age, $R_\mathrm{200}/V_\mathrm{200}\simeq 1.45\,\mathrm{Gyr}$, which corresponds to the dynamical timescale of MW-like host haloes. 
For the most massive host halo, No. 1, the number fractions of the first infalling and multiple orbiting DMSHs are 54\% and 46\%, respectively. 
The number of first infalling DMSHs is slightly higher, therefore, the violent collision frequencies of first infalling and multiple orbiting DMSHs are also higher for first infalling DMSHs, as $76\,\mathrm{Gyr^{-1}}$ and $64\,\mathrm{Gyr^{-1}}$, respectively. As a result, two types of DMSHs have caused a comparable collision frequency, but a detailed investigation of the time evolution of DMSH collision frequency and the dependency of the accretion time of DMSH into the host halo could be important for understanding galaxy evolution.

In this study, we estimate the collision frequency under idealised conditions, neglecting the mass evolution of both the host halo and the DMSHs. As this analysis is based on idealised conditions, it is intended as a foundation for subsequent, more realistic and detailed analyses.


\bibliographystyle{elsarticle-harv} 
\bibliography{Library}






\end{document}